\documentclass[12pt]{iopart}
\usepackage{graphicx}
\newcommand{\beq}{\begin{equation}}
\newcommand{\eeq}{\end{equation}}
\newcommand{\bea}{\begin{eqnarray}}
\newcommand{\eea}{\end{eqnarray}}
\renewcommand{\a}{\alpha}
\newcommand{\abs}[1]{\vert#1\vert}

\renewcommand{\b}{{\rm b}}
\renewcommand{\d}{{\rm d}}

\renewcommand{\e}{{\rm e}}
\newcommand{\eps}{\varepsilon}
\newcommand{\eq}{{\rm eq}}

\renewcommand{\lim}{{\rm lim}}
\newcommand{\lin}{{\rm lin}}
\newcommand{\m}{{\rm m}}
\newcommand{\mean}[1]{\langle#1\rangle}
\newcommand{\s}{\sigma}
\newcommand{\prob}{{\rm Prob}}
\newcommand{\thr}{{\rm th}}
\newcommand{\tot}{{\rm tot}}

\newcommand{\D}{\Delta}
\newcommand{\F}{{\cal F}}
\newcommand{\R}{{\rm R}}
\newcommand{\g}{\gamma}
\newcommand{\B}{{\rm B}}
\newcommand{\dy}{{\rm dyn}}
\renewcommand{\flat}{{\rm flat}}
\renewcommand{\frac}[2]{\displaystyle{\displaystyle#1\over\displaystyle#2}}
\newcommand{\h}{h}

\newcommand{\sign}{\mathop{\rm sign}\nolimits}
\newcommand{\xidy}{\xi_\dy}

\newcommand{\C}{{\cal C}}
\newcommand{\N}{{\cal N}}

\newcommand{\T}{\Gamma}

\newcommand{\n}{{\bf n}}

\eqnobysec

\begin{document}
\title[Competition and cooperation]
{Competition and cooperation:\\ aspects of  dynamics in sandpiles}
\author{Anita Mehta\dag,  J M Luck\S,  J. M. Berg\S\S  and G C Barker\ddag }

\address{\dag\ S N Bose National Centre for Basic Sciences, Block JD, Sector 3,
Salt Lake, Calcutta 700098, India}

\address{\S\ Service de Physique Th\'eorique\footnote{URA 2306 of CNRS},
CEA Saclay, 91191~Gif-sur-Yvette cedex, France}

\address{\S\S\ Universitat zu Koln, Zulpicher Strasse 77, 
Institut fur Theoretische Physik, D-50937 Koln, Germany}
 
\address{\ddag\ Institute of Food Research, Colney Lane, Norwich NR4 7UA, UK}

\begin{abstract}

In this article, we review some of our approaches to 
granular dynamics, now well known~\cite{book} to consist of both
fast and slow relaxational processes.
In the first case, grains typically compete
with each other, while in the second, they cooperate. A typical result of
{\it cooperation} is the formation of stable bridges, signatures of 
spatiotemporal inhomogeneities; we review their geometrical characteristics
and compare theoretical results with those of independent simulations.
{\it Cooperative} excitations due to
 local density fluctuations
are also responsible for
relaxation at the angle of repose; 
the {\it competition} between these fluctuations and external
driving forces, can, on the other hand, result
in a (rare) collapse of the sandpile to the horizontal.
Both these features are present in a theory reviewed here.
An arena where the effects of cooperation
versus competition are felt most keenly is granular compaction; we review
here a random graph model, where three-spin 
interactions are used to model compaction under tapping. The compaction
curve shows distinct regions where 'fast' and 'slow' dynamics apply,
separated by what we have called the {\it single-particle relaxation
threshold}. In the final section of this paper, we
explore the effect of shape -- jagged vs.~regular -- on the compaction
of packings near their jamming limit. One of our major
results is an entropic landscape that, while microscopically rough,
manifests {\it Edwards' flatness} at a macroscopic level. Another major
result is that of surface intermittency under low-intensity shaking.
\end{abstract}
\eads{\mailto{anita@bose.res.in},
\mailto{barker@bbsrc.ac.uk},
\mailto{berg@thp.uni-koeln.de},
\mailto{luck@spht.saclay.cea.fr}}
\maketitle

\section{INTRODUCTION}
Matter in the jammed state has become a focus of interest for physicists
in recent years.Two prime examples of this in the context of  natural systems
 are  glasses \cite{spinglass,glassyrefs} and densely packed granular media \cite{granmat}; while the mechanisms of
 jamming in each case show strong similarities, the ineffectiveness of temperature as a dynamical motor in granular media leads to vastly more surprising
effects. A direct consequence of such athermal behaviour
in sandpiles is the stable formation of
cooperative structures such as bridges \cite{jstat1}, or indeed the very
 existence \cite{br} of an angle
of repose \cite{jstat2}; neither would be possible in the presence of Brownian motion.
We discuss our studies of these two effects in the first two sections of this
article. The last two sections, with their focus on jamming, 
unify aspects of glasses and granular media; one of them uses
random graphs to illustrate competitive and cooperative effects
in granular compaction \cite{johannes}. The other concerns itself mainly
with the fast dynamics in the boundary layer of a granular column,
as the jamming limit is approached; it makes clear how asymmetric
grains can orient themselves suitably so as to waste less space,
when compelled so to do \cite{column}.

\section{On bridges in sandpiles - an overarching scenario}

The athermal nature of granular media results in the following fact:
 all granular dynamics
is the result of external stimuli. These result in grains
competing with each other to fall under gravity to a point
of stability; when instead, the process is one of cooperation
so that two or more grains fall together to rest on the substrate
with mutual support, bridges~\cite{jstat1} are formed.
These can
be stable for arbitrarily long times, since the Brownian motion
that would dissolve them away in a liquid is absent in sandpiles
-- grains are simply too large for the ambient temperature
to have any effect.
As a result, bridges can affect
the ensuing dynamics of the sandpile; a major mechanism of compaction is
the gradual collapse of long-lived bridges in weakly
vibrated granular media, resulting
in the disappearance of the voids that were earlier enclosed~\cite{usbr}.
Bridges are also responsible for jamming in granular processes,
for example, as grains flow out of a hopper~\cite{br}.

We first define a bridge in more quantitative terms.
Consider a stable packing of hard spheres under gravity, in three dimensions.
Each particle typically rests on three others which stabilise it,
in the sense that downward motion is impeded.
{\it A bridge is a configuration of particles in which
the three-point stability conditions of two or more particles
are linked; that is, two or more particles are mutually stabilised}.
Bridges thus cannot be formed sequentially, but are ubiquitous
in generic powders.
While it is impossible to determine bridge distributions
uniquely from a distribution of particle positions,
we are able via our algorithm to obtain 
the most likely positions of bridges 
in a given scenario~\cite{jstat1}.

\begin{figure}[htb]
\begin{center}
\includegraphics[angle=0,width=.38\linewidth]{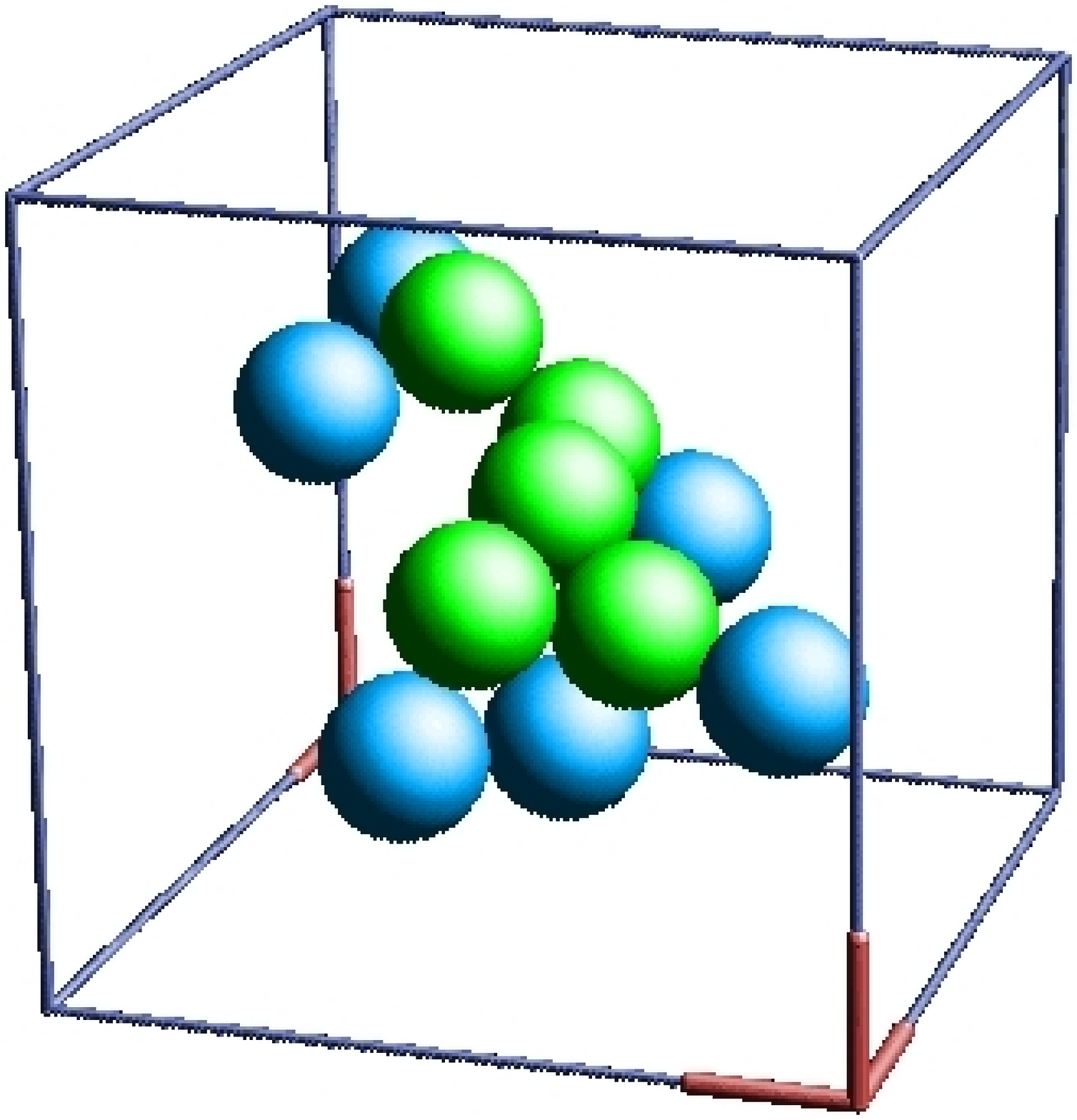}
\includegraphics[angle=0,width=.38\linewidth]{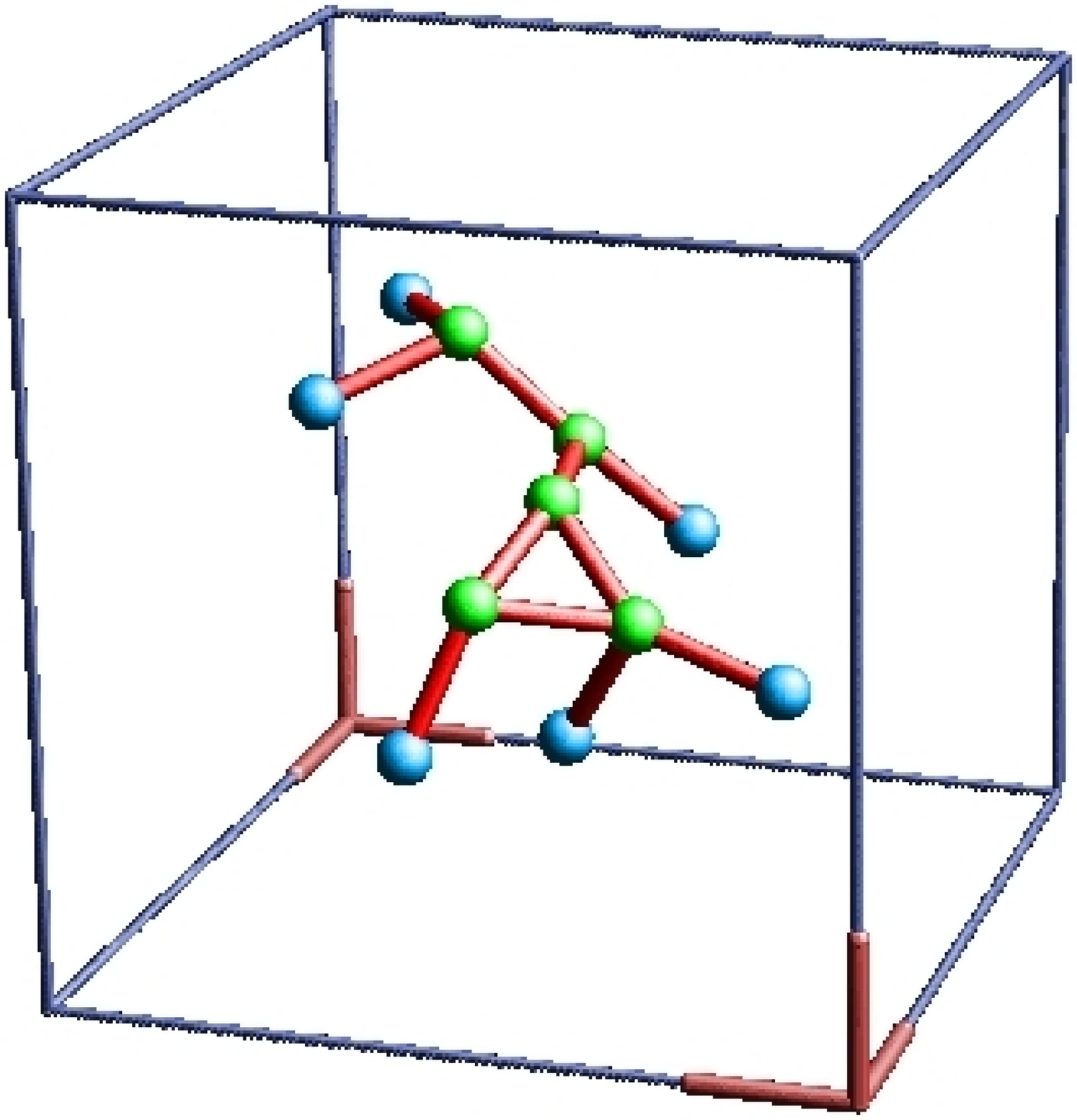}
\caption{\small
A five particle {\it complex bridge}, with six base particles (left),
and the corresponding contact network (right).
Thus $n=5$ and $n_\b=6<5+2$.}
\label{fig1}
\end{center}
\end{figure}

\begin{figure}[htb]
\begin{center}
\includegraphics[angle=0,width=.38\linewidth]{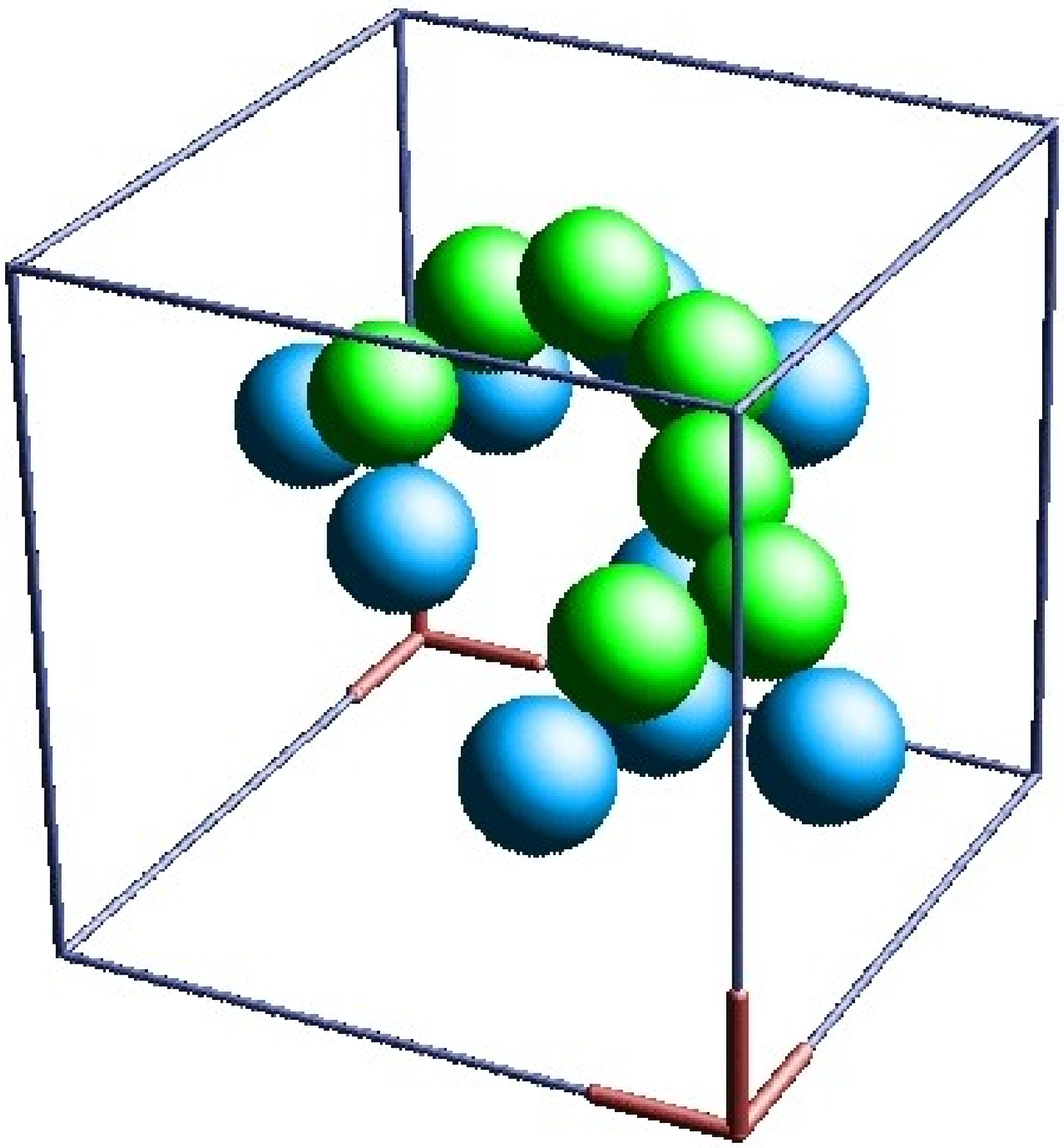}
\includegraphics[angle=0,width=.4\linewidth]{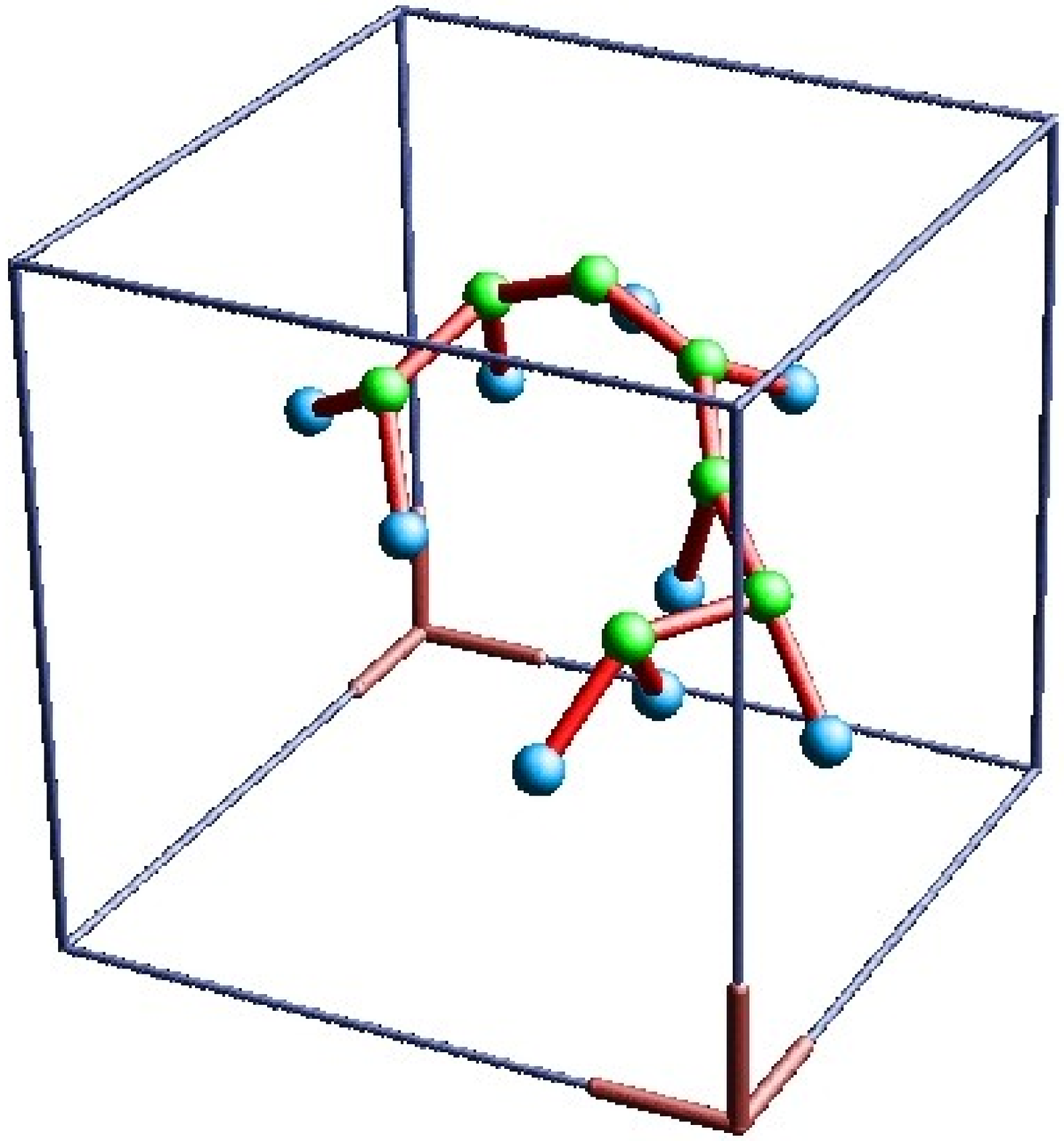}
\caption{\small
A seven particle {\it linear bridge} with nine base particles (left),
and the corresponding contact network (right).
Thus $n=7$ and $n_\b=9=7+2$.}
\label{fig2}
\end{center}
\end{figure}

We now distinguish between {\it linear} and {\it complex} bridges
via a comparison of Figures~\ref{fig1} and~\ref{fig2}.
Figure~\ref{fig1} illustrates a {\it complex} bridge,
i.e., a mutually stabilised cluster of five particles (shown in green),
where the stability is provided by six stable base particles (shown in blue).
Of course the whole is embedded in a stable network of grains within
the sandpile.
Also shown is the network of contacts for the particles in the bridge:
we see clearly that three of the particles
each have two mutual stabilisations.
Figure~\ref{fig2} illustrates a seven particle linear bridge with nine
base particles.
This is an example of a {\it linear} bridge.
The contact network shows that this bridge
has a simpler topology than that in Figure~\ref{fig1}.
Here, all of the mutually stabilised particles are in
sequence, as in a string.
A linear bridge made of $n$ particles therefore always rests
on $n_\b=n+2$ base particles.  For
a complex bridge of size $n$,
the number of base particles is reduced
($n_\b<n+2$),
because of the presence of {\it loops} in their contact networks.

An important point to note is that bridges can only be formed sustainably
in the presence of friction; the mutual stabilisations needed would
be unstable otherwise! Although our Monte Carlo
simulations (described below)  do not contain friction
explicitly, our configurations indirectly include this:
in particular, the
 coordination numbers lie in a range consistent
with the presence of friction~\cite{usbr,samfr,silbert}.

\subsection{Simulation details}

We have examined bridge structures in hard assemblies that are
generated by a non-sequential
restructuring algorithm~\cite{usbr},
whose main modelling ingredients involve {\it stochastic} grain displacements
and {\it collective} relaxation from them.

This algorithm restructures a stable
hard sphere deposit in three distinct stages.

\begin{itemize}

\item[(1)]
The granular assembly is dilated in a vertical direction (with free volume
being introduced homogeneously throughout the system), and each
particle is given a random horizontal displacement; this models
the dilation phase of a vibrated granular medium.

\item[(2)]
The assembly is compressed in a uniaxial external field representing gravity,
using a low-temperature Monte Carlo process.

\item[(3)]
Individual spheres in the assembly
 are stabilised using a steepest descent `drop and roll' dynamics to
find  local potential energy minima.

\end{itemize}
Steps (2) and (3) model the quench phase of the vibration,
where particles relax to locally stable positions in the presence of gravity.
Crucially, during the third phase, the spheres
are able to roll in contact with others; {\it mutual stabilisations} are thus
allowed to arise, mimicking collective effects.
The final configuration has
a well-defined  contact network where each sphere is supported
 by a uniquely defined set
of three other spheres.

The simulation method recalled above
builds a sequence of static packings.
Each new packing is built from its predecessor by a random process and
the sequence achieves a steady state, where
structural descriptors such as the
mean packing fraction and the mean coordination number
fluctuate about well-defined mean values.
The steady-state mean volume fraction $\Phi$ typically evolves
to values 
in the range $\Phi\sim 0.55\hbox{ -- } 0.61$, depending on the shaking
amplitude;
the mean coordination number is always $Z\approx 4.6\pm 0.1$.
We recall that for frictionless (isostatic) packings in $d$ dimensions,
 $Z=2d$ ($=6$ for $d=3$)~\cite{donev}, while for frictional packings
the {\it minimal} coordination number is
$Z=d+1$ ($=4$ for $d=3$)~\cite{samfr}; our configurations
thus clearly correspond to 
those generated in the presence of friction. This is confirmed
by the results of molecular dynamics simulations
of sphere packings
in the limit of high friction, which yield a
 mean coordination number
slightly above
 $4.5$~\cite{silbert}.

Each of our configurations includes $N_\tot\approx2200$ particles.
Segregation is avoided by choosing monodisperse particles: a rough
base prevents ordering.
A large number of restructuring cycles is needed to
reach the steady state for a given shaking amplitude:
about 100 stable configurations (picked every 100
cycles in order to avoid correlation effects) are analysed,
corresponding to
$\Phi=0.56$ and $\Phi=0.58$.
From these configurations, and following
specific prescriptions, our algorithm identifies bridges
as clusters of mutually stabilised particles~\cite{jstat1}.

\begin{figure}[htb]
\begin{center}
\includegraphics[angle=0,width=.6\linewidth]{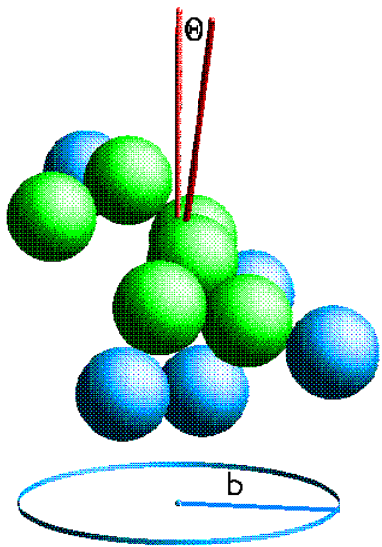}
\caption{\small
Definition of the angle $\Theta$ and the base extension $b$ of a bridge.
The
main axis makes an angle $\Theta$ with the $z$-axis; the base extension
$b$ is the projection of the radius of gyration of the bridge on the
$x$-$y$ plane.}
\label{figbig}
\end{center}
\end{figure}

Figure~\ref{figbig} illustrates two characteristic descriptors
of bridges used in this work.
The {\it main axis} of a bridge
is defined using triangulation of its base particles as follows:
triangles are constructed by choosing all possible connected triplets
of base particles, and the vector sum of their normals
is defined to be the direction of the {\it main axis} of the bridge.
The orientation angle $\Theta$ is defined as the angle
between the main axis and the $z$-axis.
The {\it base extension} $b$ is defined as the radius of gyration of
the base particles about the $z$-axis; note that this is distinct
from the radius of gyration about the main axis of the bridge.

\subsection{Bridge sizes and diameters: when does a bridge span a hole?}

In the following, we present statistics
for both linear and complex bridges.
While we recognise
that 
bridge formation 
is a collective dynamical process, we
adopt an ergodic viewpoint~\cite{ergodic} here.
Inspired by 
polymer theory~\cite{doi}, 
we visualise a linear bridge as a random chain 
which 
grows as a continuous curve, i.e. 'sequentially' in terms
of its arc length $s$. (For complex bridges,
this simplification is not possible in general
-- a direct consequence of their branched structure).
This
replacement of what is in reality a {\it collective phenomenon in time by
a random walk in space} is somewhat analogous to  the 'tube model'
of linear polymers~\cite{doi}: both are simple but efficient {\it effective}
pictures of very complex problems.

We first address the question of the length distribution
of linear bridges.
We define the length distribution $f_n$ as the probability that a linear bridge
consists of exactly $n$ spheres.
We make the simplest and the most natural assumption
that a bridge of size $n$ remains linear with some probability $p<1$
if an $(n+1)^{\rm th}$ sphere is 'added' to it: this leads to
the exponential distribution
\beq
f_n=(1-p)p^n.
\label{fexp}
\eeq

The exponential distribution above 
can also be derived
by means of a continuum approach.
Here, a linear bridge is viewed as a continuous random curve or `string',
parametrised by the arc length $s$ from one of its endpoints.
We assume also that such a bridge
disappears at a constant rate $\a$ per unit length,
either by changing from linear to complex or by collapsing.
The survival probability $S(s)$ of a linear bridge upto length~$s$ 
thus obeys the rate equation
$\dot S=-\a S$ and falls off exponentially,
according to $S(s)=\exp(-\a s)$. Consequently,
the probability distribution of the length $s$ of linear bridges
 reads $f(s)=-\dot S(s)=\a\,\exp(-\a s)$, a
 continuum analogue of~(\ref{fexp}).

This is in good accord with the results of independent simulations,
which
 exhibit an exponential decay of linear bridges of the form~(\ref{fexp}),
with $\a\approx0.99$ \cite{jstat1}, which is
 clearly seen
until $n\approx12$. Around
$n\approx8$, complex bridges begin to predominate; these
have size distributions which show a power-law decay:
\beq
f_n\sim n^{-\tau}.
\label{fpower}
\eeq
with 
  $\tau\approx2$ \cite{jstat1}.

We have also measured 
the  diameter $R_n$ of linear and complex bridges  of size $n$,
which is
 such that $R_n^2$ is the mean squared end-to-end distance.
Our data on diameters and size distributions~\cite{jstat1}
indicate that linear bridges in three dimensions start off as {\it planar
self-avoiding walks}, which eventually
 collapse onto each other because of vibrational effects; on the other hand,
complex bridges look like
 {\it $3d$  percolation clusters}.

Another issue of interest to us is the
jamming potential of a bridge. A measure of this, in the case
of a linear  bridge, is its
the base extension~$b$ (see Figure~\ref{figbig}); this is
the horizontal projection of the 'span' of the bridge.
Our simulation results \cite{jstat1} indicate that {\it three-dimensional bridges of a given length have a fairly characteristic horizontal extension}, making it relatively
easy to predict whether or not they
would 'jam' a given hole. 

\begin{figure}[htb]
\begin{center}
\vskip 10pt
\includegraphics[angle=0,width=.6\linewidth]{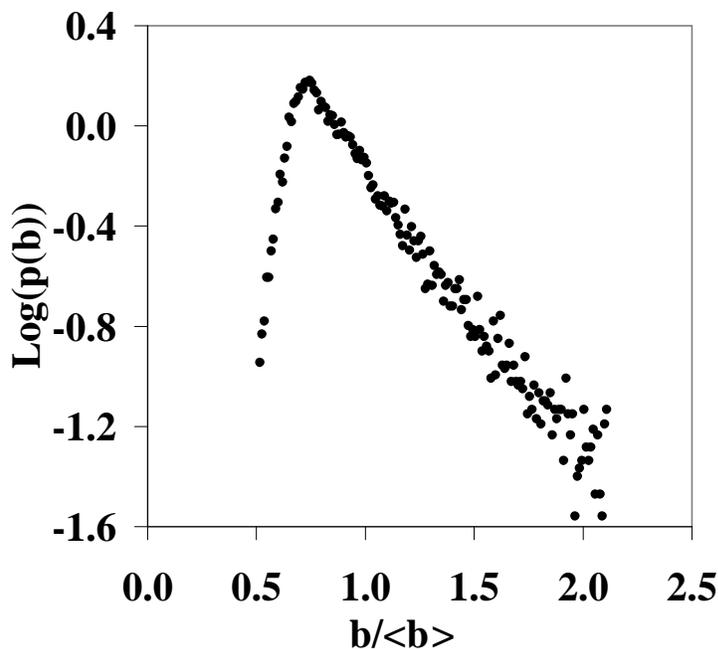}
\caption{\small
Distribution of base extensions of bridges, for $\Phi=0.58$.
The logarithm of the normalised probability distribution is plotted
as a function of the normalised variable $b/\mean{b}$,
where~$\mean{b}$ is the mean extension of bridge bases.}
\label{fig8}
\end{center}
\end{figure}

In order to compare our simulations with experiment~\cite{pak, mueth},
we plot in Figure~\ref{fig8}  the logarithm
of the probability distribution of base extensions $p(b)$
against the (normalised) base extension
$b/\mean{b}$. This figure emphasises the exponential tail of the distribution function,
and also shows that bridges with small base extensions are unfavoured.
We note that this long tail
is characteristic of three-dimensional
experiments on force chains in granular media \cite{mueth}.
The sharp drop at the origin as well as the long
tail in  Figure~\ref{fig8}, are observed in
normal force distributions obtained via
 molecular dynamics simulations of particle packings~\cite{ohern},
in the limit of strong
deformations.
Realising that the measured
forces propagate through chains of particles,
we use this similarity to suggest that {\it bridges are
really just long-lived force chains}, which have survived
despite strong deformations.
We suggest also that
with the current availability of 3D
visualisation techniques such as NMR~\cite{fukushima},
bridge configurations might be an easily measurable
and effective tool to probe inhomogeneities
 in shaken sand.

\subsection{Turning over at the top; how linear bridges form domes}

Recall that
a linear bridge is modelled as a continuous curve,
parametrised by its arc length $s$.
We here focus on its most important degree of freedom,
the tilt with respect to the horizontal;  the azimuthal degree
of freedom is neglected.
Accordingly, we define the local or {\it link} angle $\theta(s)$
between the direction of the tangent to the bridge at point~$s$
and the horizontal,
and the {\it mean angle} made by the bridge from its origin up to point~$s$, also with the horizontal:
\beq
\Theta(s)=\frac{1}{s}\int_0^s\theta(u)\,\d u.
\label{Thdef}
\eeq
The local angle $\theta(s)$ so defined may be either
positive or negative; it can even change sign
along the random curve which represents a linear bridge.
Of course, the orientation angle $\Theta$ measured in our numerical simulations
is positive by construction,
being defined  as the angle between the main 
bridge axis
and the $z$-axis (see Figure \ref{figbig}). (Note that by simple geometry,
this 'zenith angle' made by the {\it bridge axis} with the {\it vertical},
equals the mean angle
made by the {\it basal plane} of the bridge with the {\it horizontal}).

Our simulations show that
the mean angle $\Theta(s)$ typically becomes smaller and smaller
as the length $s$ of the bridge increases.
Small linear bridges are almost never flat~\cite{jstat1}; as they get
longer, assuming that they still stay linear, they get `weighed down', arching
over as at the mouth of a hopper~\cite{br}.
Thus, in addition to our earlier claim that long
linear bridges are rare, we claim further here that
(if and) when they exist, they typically have flat bases, becoming `domes'.

We use these insights to write down equations
to investigate the angular distribution of linear bridges.
These couple the evolution of the local angle $\theta(s)$
with local density fluctuations $\phi(s)$ at point~$s$ 
(with $'$ denoting a derivative with respect to $s$):
\bea
&&\theta'=-a\theta-b\phi^2+\D_1\eta_1(s),
\label{cou1}\\
&&\phi'=-c\phi+\D_2\eta_2(s).
\label{cou2}
\eea
The effects of vibration on each of $\theta$ and $\phi$ are represented by
two independent white noises $\eta_1(s)$, $\eta_2(s)$, such that
\beq
\mean{\eta_i(s)\eta_j(s')}=2\,\delta_{ij}\,\delta(s-s'),
\label{white}
\eeq
whereas the parameters $a$,~...,~$\D_2$ are assumed to be constant.

The phenomenology behind the above equations is the following:
the evolution of $\theta(s)$ is caused, in our effective picture,
by the {\it sequential} addition  
of particles to the bridge at its ends.
The fluctuations
of local density $\phi$ at a point $s$
are caused by {\it collective} particle motion~\cite{mln}.
The
first terms on the right-hand side of~(\ref{cou1}),~(\ref{cou2})
say that neither $\theta$ nor $\phi$ is allowed to be arbitrarily large.
Their coupling via the second
term in~(\ref{cou1}) arises as follows: if there are 
density fluctuations $\phi^2$ of large magnitude at the tip
of a bridge, 
 these will,
to a first approximation,
 `weigh the bridge down', i.e.,
decrease the angle $\theta$ locally.

Reasoning as above, we therefore anticipate that for low-intensity vibrations
and stable bridges, both density fluctuations $\phi(s)$
and link angles $\theta(s)$ will be small.
Accordingly, we linearise~(\ref{cou1}),
obtaining thus an Ornstein-Uhlenbeck equation:
\beq
\theta'=-a\theta+\D_1\eta_1(s).
\label{ou}
\eeq
Let us make the additional assumption that the initial angle $\theta_0$,
i.e., that observed for very small bridges,
is itself Gaussian with variance $\s_0^2=\mean{\theta_0^2}$.
The angle $\theta(s)$ is then a Gaussian process with zero mean
for any value of the length $s$.
Its correlation function can be easily evaluated to be~\cite{ou}:
\beq
\mean{\theta(s)\theta(s')}
=\s_\eq^2\,\e^{-a\abs{s-s'}}+(\s_0^2-\s_\eq^2)\e^{-a(s+s')}.
\label{thcor}
\eeq
It follows from this that the variance of the link angle
is:
\beq
\mean{\theta^2}(s)=\s_\eq^2+(\s_0^2-\s_\eq^2)\e^{-2as},
\eeq
We see from the above that orientation correlations
decay with a characteristic length given by
$\xi=1/a$; also, in the limit of an infinite
bridge, the  variance $\mean{\theta^2}$ relaxes
to $\s_\eq^2=\frac{\D_1^2}{a}$~\cite{jstat1}. Thus,
as the chain gets longer, the variance of the link angle relaxes
from its initial value of
$\s_0^2$ (i.e. that for the initial link) to $\s_\eq^2$ for infinitely
long chains.

Given the above, it can be shown that the mean angle $\Theta(s)$
will also have a Gaussian distribution.
Its variance can be derived by inserting~(\ref{thcor}) 
into~(\ref{Thdef}):
\beq
\mean{\Theta^2}(s)=2\s_\eq^2\,\frac{as-1+\e^{-as}}{a^2s^2}
+(\s_0^2-\s_\eq^2)\frac{(1-\e^{-as})^2}{a^2s^2}.
\label{Thvar}
\eeq
The asymptotic result
\beq
\mean{\Theta^2}(s)\approx\frac{2\s_\eq^2}{as}\approx\frac{2\D_1^2}{a^2s},
\label{thas}
\eeq
confirms our earlier statement that {\it the longest bridges
form domes}, i.e. they have bases that are almost flat.
Each such bridge can be viewed as
consisting of a large number $as=s/\xi\gg1$ of
independent `blobs' of length $\xi$; this result
suggests, yet again, strong analogies between linear bridges and
linear polymers \cite{doi}.

The result~(\ref{thas}) has another interpretation.
As $\Theta(s)$ is small with high probability for a very long bridge,
its extension in the vertical direction reads approximately
\beq
Z=z(s)-z(0)\approx s\,\Theta(s),
\label{zed}
\eeq
so that $\mean{Z^2}\approx s^2\mean{\Theta^2}(s)\approx2(\D_1/a)^2\,s$.
Switching back to a discrete picture of an $n$-link chain, we have
\beq
Z_n\sim n^{1/2}.
\eeq
The vertical extension of a linear bridge is thus found to grow with the usual
random-walk exponent $1/2$, in agreement with
experiments on two-dimensional arches~\cite{pak}.
Our observations on horizontal extensions of three-dimensional
bridges have yielded~\cite{jstat1} a
non-trivial exponent $\nu_\lin\approx0.66$.
Putting all of this together, our results predict that
{\it long linear bridges are domelike; also, they are
vertically diffusive but horizontally superdiffusive}.
Evidently, jamming in a three-dimensional hopper would be caused
by the planar projection of such a {\it dome}.

\begin{figure}[htb]
\begin{center}
\includegraphics[angle=90,width=.57\linewidth]{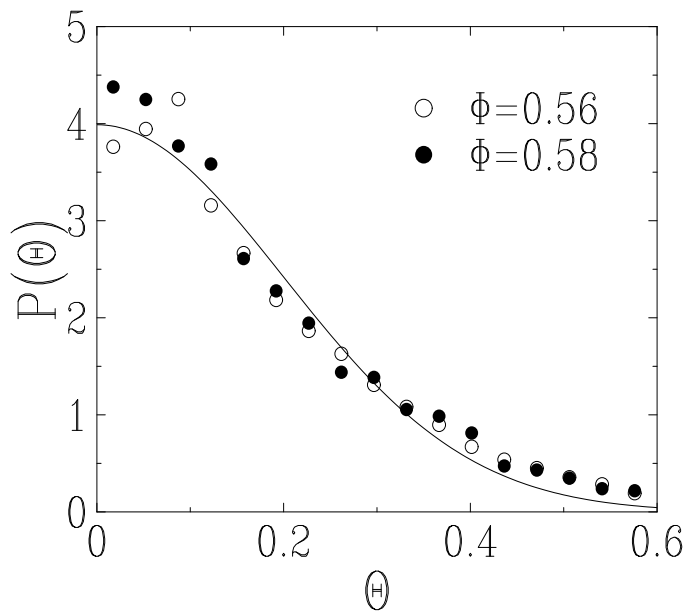}
\caption{\small
Plot of the normalised distribution of the mean angle $\Theta$
(in radians) of linear bridges of size $n=4$, for both volume fractions.
The $\sin\Theta$ Jacobian has been duly divided out,
explaining thus the larger statistical errors at small angles.
Full lines: common fit to (half) a Gaussian law.}
\label{figgauss}
\end{center}
\end{figure}

\begin{figure}[htb]
\begin{center}
\includegraphics[angle=90,width=.50\linewidth]{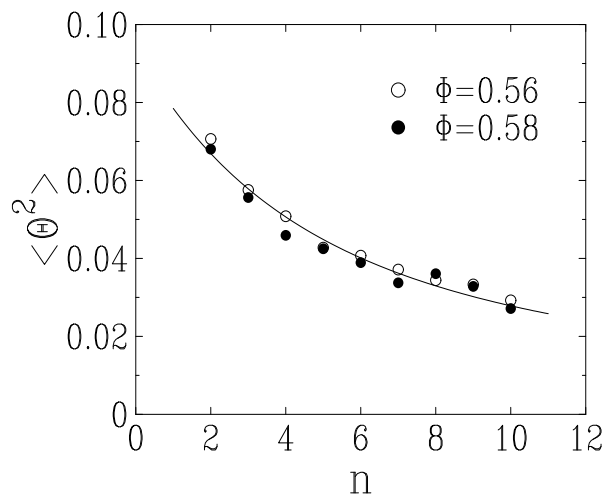}
\caption{\small
Plot of the variance of the mean angle of a linear bridge,
against size~$n$, for both volume fractions.
Full line: common fit to the first (stationary) term of~(\ref{Thvar}),
yielding $\s_\eq^2=0.093$ and $a=0.55$.
The `transient' effects of the second term of~(\ref{Thvar}) are invisible
with the present accuracy.}
\label{figvar}
\end{center}
\end{figure}

We now compare the results of this simple theory with data on bridge
structures obtained
from {\it independent} numerical simulations of shaken hard sphere packings \cite{usbr}.
Figure~\ref{figgauss}
confirms that the mean angle is Gaussian to a good approximation, while
Figure~\ref{figvar} shows the measured size dependence
of the variance $\mean{\Theta^2}(s)$.
The numerical data are found to agree well with a common fit
to the first (stationary) term of~(\ref{Thvar}) -
the `transient' effects of the second term of~(\ref{Thvar}) are too small
to be significant at our present accuracy.
We thus conclude that our simple 
theory captures the principal structural features of linear bridges.

\subsection{Discussion}

We end this section with the following remarks.
First, more subtle effects,
including the effects of transients via the second term of~(\ref{Thvar}),
and the dependence of the parameters $\s_\eq^2$ and~$a$
on the packing fraction $\Phi$,
are deserving of further investigation.
Second, we might expect that
 with increasing density $\Phi$, branched structures
would become more and more common; linear bridge
formation, with its
'sequential' progressive attachment of independent blobs would then
become more and more rare. Our theory
 should therefore cease to hold
at a limit packing fraction $\Phi_\lim$,
which is qualitatively reminiscent of the  single-particle
relaxation threshold density~\cite{johannes} (see section $3$).
Finally,
 our investigations suggest that {\it long-lived bridges
are natural indicators of sustained inhomogeneities in granular systems}.

\section{On angles of repose: bistability and collapse}

Typically, the faces of a sandpile are inclined
at a finite angle to the horizontal.
This is the so-called 'angle of repose'~$\theta_\R$: in practice,
 it can take a range of values before spontaneous
flow occurs, i.e., the sandpile becomes unstable to further deposition.
The limiting
value of this pre-avalanching angle is known as the {\it maximal
angle of stability} $\theta_\m$~\cite{br}.
Also, as a result of their athermal nature, sandpiles
are strongly hysteretic; this results
in {\it bistability} at the angle of repose
~\cite{bistability,daerr},
 such that
 a sandpile can either be stable. or in motion,
at any angle~$\theta$ such that $\theta_\R<\theta<\theta_\m$.
Notwithstanding the above, it is possible for a sandpile
to undergo spontaneous collapse
to the horizontal; this is, in general, a rare event.
We propose a theoretical explanation \cite{jstat2} below for both
 bistability at, and collapse through, the angle of repose
via the coupling of fast and slow relaxational modes in a sandpile~\cite{book}.

\subsection{Coupled nonlinear equations: dilatancy vs. the angle of repose}

Our basic picture is that fluctuations of local density
are the collective excitations responsible for stabilising the angle of repose,
and for giving it its characteristic width,
\beq
\delta\theta_\B=\theta_\m-\theta_\R,
\label{ba}
\eeq
known as the Bagnold angle~\cite{bagnold}.
Such
density fluctuations may arise from, for instance, shape
effects~\cite{column} or friction~\cite{br,edwards98};
they are the manifestation in our model of 
{\it Reynolds dilatancy}~\cite{reynolds}.

The dynamics of the angle of repose $\theta(t)$ and of the density
fluctuations $\phi(t)$ are described \cite{jstat2} by the following
 stochastic equations, which couple their time derivatives $\dot\theta$
and $\dot\phi$:
\bea
&&\dot\theta=-a\theta+b\phi^2+\D_1\,\eta_1(t),
\label{dy1}\\
&&\dot\phi=-c\phi+\D_2\,\eta_2(t).
\label{dy2}
\eea
The parameters $a$,~...,~$\D_2$ are phenomenological constants,
while $\eta_1(t)$, $\eta_2(t)$ are two independent white noises such that
\beq
\mean{\eta_i(t)\eta_j(t')}=2\,\delta_{ij}\,\delta(t-t').
\label{white1}
\eeq

The first terms in~(\ref{dy1}) and~(\ref{dy2}) suggest that
neither the angle of repose nor the dilatancy is allowed
to be arbitrarily large for a stable system.
The second term in~(\ref{dy1}) affirms that dilatancy underlies the phenomenon
of the angle of repose;  in the absence of noise,
density fluctuations {\it constitute} this angle. The
 term proportional to $\phi^2$ is written on symmetry grounds,
since the 
 the magnitude (rather than the sign)
of density fluctuations should determine the width
of the angle of repose.
The noise in~(\ref{dy1}) represents external vibration,
while that in~(\ref{dy2}) embodies  Edwards'
compactivity~\cite{ergodic}, being related
to purely density-driven effects. We note that these equations
bear more than a passing resemblance to those in the previous
section on orientational statistics of bridges: the underlying
reason for this similarity is the idea~\cite{jstat1, jstat2} that bridges form by initially
aligning themselves at the angle of repose in a sandpile.

Examining the above equations, we quickly distinguish two regimes.
When the material is weakly dilatant ($c\gg a$),
so that density fluctuations decay quickly to zero (and hence can be neglected),
the angle of repose $\theta(t)$ 
 relaxes {\it exponentially fast} to an equilibrium state, whose variance
\beq
\theta_\eq^2=\frac{\D_1^2}{a}
\eeq
is just the zero-dilatancy variance of $\theta$.
The opposite limit, where $c\ll a$, 
and  density fluctuations are long-lived,
will be our regime of interest here.
When, additionally, $\D_1$ is small, the angle
of repose has a {\it slow} dynamics reflective of the slowly evolving
density fluctuations. 
These conditions can be written more precisely as
\beq
\g\ll1,\qquad\eps\ll1,
\label{regime}
\eeq
in terms of two dimensionless parameters (see~(\ref{thesca}):
\beq
\g=\frac{c}{a},\qquad
\eps=\frac{ac^2\D_1^2}{b^2\D_2^4}=\frac{\theta_\eq^2}{\theta_\R^2}.
\label{regdef}
\eeq

The parameter $\g$, which sets the separation of the 
fast and slow timescales, is an inverse measure of
 {\it dilatancy} in the granular medium; small
 values of this imply a granular medium
that is 'stiff' to deformation, resulting
from the persistence of density fluctuations. The parameter $\eps$
measures the ratio of fluctuations about the (zero-dilatancy)
 angle of repose
to its full value in the presence of density fluctuations: from this we can
already infer that it is a {\it measure of the ratio of the external
vibrations to density-driven effects}, which are explicitly contained
in the ratio
 ($\frac{\D_1^2}{\D_2^4}$). Realising that {\it external vibrations
and density/compactivity respectively drive fast and slow
dynamical processes in a granular system}, we see that a quantity
which measures their ratio has all the characteristics
of an effective temperature~\cite{book} in the slow dynamical
regime of interest to us here. This temperature-like aspect
will become much more vivid subsequently, when we discuss the issue
of sandpile collapse.

To recapitulate: the
 regime~(\ref{regime}) that we will discuss below
 is characterised as
 {\it low-temperature and strongly dilatant}, 
governed as it is by the {\it slow dynamics} of density fluctuations.

\subsection{Bistability within $\delta\theta_\B$ : how dilatancy 'fattens' the angle of repose}

Suppose that a sandpile is created in regime~(\ref{regime}) with very large
initial values for the angle $\theta_0$ and dilatancy $\phi_0$.
In the  initial transient stages, the noises have
negligible effect and the decay is
governed by the deterministic parts
of~(\ref{dy1}) and~(\ref{dy2}):
\bea
&&\theta(t)=(\theta_0-\theta_\m)\e^{-at}+\theta_\m\,\e^{-2ct},\\
&&\phi(t)=\phi_0\,\e^{-ct},
\label{decay}
\eea
with
\beq
\theta_\m\approx\frac{b\,\phi_0^2}{a}.
\label{thmax}
\eeq
Thus,
density fluctuations $\phi(t)$ relax ex\-po\-nen\-tially,
while the trajectory $\theta(t)$ has two separate modes of relaxation. First,
there is
a fast (inertial) decay in $\theta(t)\approx\theta_0\,\e^{-at}$,
until $\theta(t)$ is of the order of $\theta_\m$; this is followed by
a slow (collective) decay in $\theta(t)\approx\theta_\m\,\e^{-2ct}$.
When $\phi(t)$ and $\theta(t)$
are small enough [i.e., $\phi(t)\sim\phi_\eq$
and $\theta(t)\sim\theta_\R$, cf.~(\ref{phisca}) and~(\ref{thesca})]
for the noises to have an appreciable effect, the above analysis
is no longer valid.
The system then reaches the equilibrium state
of the full non-linear stochastic process
represented by~(\ref{dy1}) and~(\ref{dy2}), a full analytical
solution of which is presented in \cite{jstat2}.

In order to get a feeling for the more qualitative features
of the equilibrium state,
we note first that the equilibrium variance of
 $\phi(t)$ is: 
\beq
\phi_\eq^2=\frac{\D_2^2}{c}.
\label{phisca}
\eeq
We see next that to a good approximation,
the angle $\theta$ adapts instantaneously to the dynamics of $\phi(t)$
in regime~(\ref{regime}):
\beq
\theta(t)\approx\frac{b\,\phi(t)^2}{a}.
\label{thinst}
\eeq
The two above statements together imply that
the distribution of the angle~$\theta(t)$ is approximately
that of the square of a Gaussian variable.
The {\it typically observed} angle of repose $\theta_\R$
is the time-averaged value
\beq
\theta_\R=\mean{\theta}_\eq=\frac{b\,\phi_\eq^2}{a}=\frac{b\D_2^2}{ac}.
\label{thesca}
\eeq
Equation~(\ref{thinst}) then reads
\beq
\theta(t)\approx\theta_\R\,\frac{\phi(t)^2}{\phi_\eq^2}.
\label{thred}
\eeq
Equation~(\ref{thred}) entirely explains
the physics behind the multivalued and history-dependent
nature of the angle of repose~\cite{sidrevmodphys, sidprl}.
 Its instantaneous value depends directly
on the instantaneous value of the dilatancy; its maximal (stable) value
$\theta_\m$ is noise-independent
[cf.~(\ref{thmax})] and depends only on the maximal
value of dilatancy that a given material can sustain stably~\cite{br}.
Sandpiles constructed above this
will first decay quickly
to it; they will then decay more slowly to a `typical'
angle of repose~$\theta_\R$.
The ratio of these angles is given by
\beq
\frac{\theta_\m}{\theta_\R}=\frac{\phi_0^2}{\phi_\eq^2},
\label{thth}
\eeq
so that $\theta_\m\gg\theta_\R$ for $\phi_0\gg\phi_\eq$.
Within the Bagnold angle $\delta\theta_\B$, 
(i.e. for sandpile inclinations which lie in the range
$\theta_\R<\theta<\theta_\m$), our simple theory also demonstrates
the presence 
of {\it bistability}.
Thus, sandpiles submitted to low noise are stable
in this range of angles (at least for long times $\sim 1/c$); on the other 
hand, sandpiles submitted to  high  noise (such that the effects
of dilatancy become negligible in~(\ref{dy1}))
 continue to decay rapidly in this range of angles,
becoming nearly horizontal at short times $\sim 1/a$.

Our conclusions are that bistability at the angle
of repose is a natural consequence of applied noise
(tilt~\cite{bistability} or vibration) in granular systems.
For sandpile inclinations $\theta$ within the range $\delta\theta_\B$, 
sandpile history is all-important: depending on this,
a sandpile can either be at rest or in motion
at the {\it same} angle of repose.

\subsection{When sandpiles collapse: rare events, activated processes
and the topology of rough
landscapes}

When sandpiles are subjected to low noise for a sufficiently
long time, they can collapse \cite{book}, such that the
angle $\theta(t)$ vanishes.
Such an event is expected to be very rare in the regime~(\ref{regime});
in fact it occurs only if the noise~$\eta_1(t)$
in~(\ref{dy1}) is sufficiently negative
for sufficiently long to compensate for the strictly positive term~$b\phi^2$.
It can be shown~\cite{jstat2} that the
equilibrium probability for $\theta$ to be negative,
$\Pi=\prob(\theta<0)$,
scales throughout regime~(\ref{regime}) as:
\beq
\Pi\approx\frac{(2\eps)^{1/4}}{\Gamma(1/4)}\,\F(\zeta),\qquad
\zeta=\frac{\g}{\eps^{1/2}}=\frac{b\D_2^2}{a^{3/2}\D_1}
\label{pitext}
\eeq
The scaling function $\F(\zeta)$
decays \cite{jstat2} mo\-no\-to\-ni\-cally from $\F(0)=1$ to $\F(\infty)=0$; 
to find out when the angle of repose first crosses zero, we should
explore the latter limit, i.e.
 the regime $\zeta\gg1$. 
Here,  the equilibrium probability
of collapse vanishes exponentially fast:
\beq
\Pi\sim\exp\left(-\frac{3}{2}\left(\frac{\g^2}{\eps}\right)^{1/3}\right).
\label{act}
\eeq
The above suggests that sandpile collapse is an
 {\it activated} process, with a {\it competition} between
 'temperature' $\eps$ and 'barrier height'  $\g^2$.
Collapse events occur at Poissonian times,
with an exponentially large characteristic time given by an Ar\-rhe\-nius~law:
\beq
\tau\sim1/\Pi
\sim\exp\left(\frac{3}{2}\left(\frac{\g^2}{\eps}\right)^{1/3}\right).
\label{arrhe}
\eeq
The stretched exponential with a fractional power of
the usual 'barrier-height-to-temperature ratio' $\g^2/\eps$ 
is suggestive of glassy dynamics \cite{glassyrefs}; it also
reinforces the idea that sandpile collapse is a {\it rare event}.

While the reader is referred to a longer paper \cite{jstat2} for 
the derivation of the stretched exponential,
the physics behind it is readily understood by means of an exact
analogy with the 
 problem of random trapping~\cite{trapping}, which we
outline below.

Consider a Brownian particle in one dimension,
diffusing (with diffusion constant $D$)
among a concentration $c$ of Poissonian traps. 
Once a trap is reached, the particle ceases to exist, so that its
 survival probability $S(t)$ 
is also the probability that it has not encountered a trap until time $t$.
Assuming a uniform distribution of starting points,
the fall-off of this probability can be estimated by
first computing the probability of finding a large region of length $L$ without
traps, and then weighting this with the probability that
a Brownian particle survives within it for a long time~$t$:
\beq
S(t)\sim\int_0^\infty\exp\left(-cL-\frac{\pi^2Dt}{L^2}\right)\,\d L.
\label{int2}
\eeq
The first exponential factor $\exp(-cL)$ is the probability that a
region of length $L$ is free of traps, whereas the second exponential factor
is the asymptotic survival probability of a Brownian particle
in such a region,  $\exp(-Dq^2t)$.
 The integral is dominated by a saddle-point at $L\approx\left(\frac{2\pi^2Dt}{c}\right)^{1/3}$, whence we recover the well-known estimate
\beq
S(t)\sim\exp\left(-\frac{3}{2}\left(2\pi^2c^2Dt\right)^{1/3}\right).
\label{sur}
\eeq

Notice the similarity in the forms of~(\ref{act}) and~(\ref{sur}); it turns
out that the steps in their derivations are identical~\cite{jstat2},
and form the basis of an exact analogy. In turn the analogy allows
us to formulate an {\it optimisation-based}
 approach to sandpile collapse, which
makes for a much more intuitive grasp of its physics.

Accordingly, let us visualise the angle $\theta$ as an 'exciton'
whose 'energy levels' are determined by the magnitude of $\theta$.
It diffuses with temperature $\eps$
in a frozen landscape of
$\phi$ (dilatancy) barriers 
of typical energy~$\g$. 
Only if it succeeds in finding
an unusually low barrier can it escape via
~(\ref{act}), to
  reach its ground state ($\theta=0$) - this of course
corresponds to sandpile collapse.
Taking the analogy a step further, we visualise the exciton
as 'flying' at a 'height' $\theta$, surrounded by $\phi$-peaks of typical
'height' $\g$ in a rough landscape.
Flying too low would cause the $\theta$ exciton to hit a $\phi$ barrier fast,
while
flying too high would cause the exciton to miss the odd low barrier. 
It turns out~\cite{jstat2} that flying at 
 $\theta\sim\eps^{1/3}$ allows the exciton to escape
via~(\ref{act})
(cf. the arguments leading to $L\sim t^{1/3}$ above).
Translating back to the scenario of sandpile angles, the above
arguments imply the following:
angles of repose that are too low
are unsustainable for any length of time, given dilatancy
effects, while angles that are too large will resist collapse.
Thus {\it optimal angles for sandpile collapse are found to scale as
$\theta\sim\eps^{1/3}$}; sandpiles with these inclinations
show a finite, if small, tendency to collapse via~(\ref{act}). 

Clearly, the frequency of collapse will depend on the topology
of the $\phi$-landscape; the form~(\ref{act}) was valid for a
landscape with Gaussian roughness~\cite{jstat2}. What if the landscape
is much rougher or smoother than this? - to answer this question,
we look at two opposite extremes of non-Gaussianness.

First, let us assume that
 density fluctuations are peaked around zero; 
typical barriers are  low, and the
 $\phi$-landscape is much flatter than Gaussian. The exciton's
escape probability ought now to be greatly increased. This is in
fact the case~\cite{jstat2}; it can be shown that in the $\g\to0$ limit,
the collapse probability scales as $\eps^{1/4}$.
Switching back to the language of sandpiles, this limit corresponds
to a nearly {\it non-dilatant material}; it results in
a `liquid-like' scenario of {\it frequent collapse},
where a finite angle of repose is hard to sustain under any
circumstances.

In the opposite limit of an extremely rough energy landscape,
where large values of $\phi$ are more frequent than in the Gaussian
distribution, one might expect the escape
probability of the $\theta$ exciton to be greatly reduced.
If, for example,
the jaggedness of the landscape is such that
 $\abs{\phi(t)}$
is always larger than some threshold $\phi_\thr$,
the stretched exponential in~(\ref{act}) reverts
(in the $\eps\ll1$ regime considered) to an Arrhenius law in its usual form:
\beq
\Pi\sim\exp\left(-\frac{(\phi_\thr/\phi_\eq)^4}{2\eps}\right).
\eeq
In the language of sandpiles, this limit corresponds to {\it strongly
dilatant} material; here, as one might expect,
sandpile collapse is even {\it more strongly inhibited} than in~(\ref{act}).  
Wet sand for example, is strongly dilatant; its angles
of repose can be far steeper than usual,          
and still resist collapse.

\subsection{Discussion}

The essence of our
theory above is that  dilatancy is responsible for the existence
of the angle of repose in a sandpile. We claim further that bistability at 
the angle of repose 
results from the difference between out-of-equilibrium
and equilibrated dilatancies. We are also
able to provide an analytical confirmation
of the following everyday observation:
{\it weakly dilatant sandpiles collapse easily, while
strongly dilatant ones bounce back}.

\section{Compaction of disordered grains in the jamming limit: 
sand on random graphs}

Granular compaction is characterised by a {\it competition between
fast and slow degrees of freedom}~\cite{book}; when voids abound, individual
grains can quickly move into these to suit their convenience.
As the jamming limit is approached, clusters of grains need to
rearrange
in order better to fill remaining partial voids: this process is necessarily
slow, and eventually leads to {\it dynamical
arrest} \cite{glassyrefs}. 

The modelling of granular compaction
has been the subject of considerable effort.
Early simulations of
shaken hard sphere packings~\cite{usbr}, carried out in close symbiosis
with experiment~\cite{sid1}, were followed by lattice-based theoretical
models~\cite{usepl, tetris}; the latter could  not, of course,
incorporate the reality of a disordered substrate. Mean-field
models~\cite{jorge} which could incorporate such disorder could not,
on the other hand, impose the finite connectivity of grains
included in Refs.~\cite{usbr, usepl, tetris}. It was to answer the need of an analytically tractable model
which incorporated
{\it finitely connected grains on fully disordered substrates}, that we
introduced~\cite{johannes} random graph models of granular compaction.

Why random graphs?  First, 
random graphs~\cite{rg} are the simplest structures with a {\it finite} number of neighbours. Finite connectivity is a key property
of grains in a sandpile with a gamut of
consequences, ranging from kinetic constraints~\cite{kobandersen} 
to cascade processes in granular compaction~\cite{johannes}.
Second, random graphs are among the simplest fully {\it disordered}  
constructs where, despite the existence of defined neighbourhoods of a site,
no global symmetries exist. Disorder is an equally key feature of granular 
matter,  even at the highest
densities; its consequences include the presence of a range
of coordination numbers~\cite{usbr} for any sandpile,
 corresponding to {\it locally
varying neighbourhoods} of individual grains, a feature
which can be
incorporated via {\it locally fluctuating connectivities}~\cite{johannes}
 in random graphs.

Having motivated our choice of random graphs as a basis,
 we proceed below to describe
 our spin model of granular compaction.

\subsection{The three-spin model: frustration, metastability and slow dynamics}
  
\begin{figure}[htb]
\begin{center}
\includegraphics[angle=0,width=.50\linewidth]{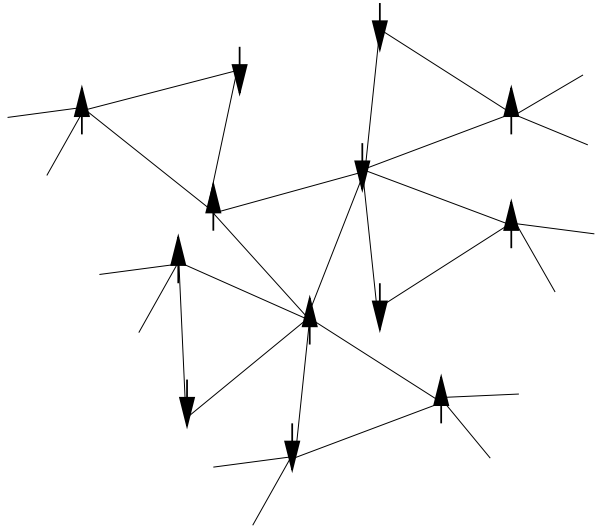}
\caption{\small
A part of a random graph with triplets of sites forming plaquettes
illustratng its local treelike structure (no planarity or geometric
sense of distance are implied).}
\label{3graph}
\end{center}
\end{figure}

The guiding factor in our choice of spin model is that it be the simplest
model with frustration, metastability and slow dynamics; we will discuss
the last two later, but remark at the outset that {\it geometrical 
frustration} is crucial to any study of granular matter. 
This concerns the fruitless {\it competition} between grains
which try - and fail - to fill voids in the jamming limit, due
either to geometric constraints on their mobility, or because
of incompatibilities in shape or size.
Our way of modelling this is via
multi-spin 
interactions on plaquettes on a random graph \cite{johannes}. 
We choose in particular
a 3-spin Hamiltonian on a 
random graph (see Fig.~\ref{3graph}) where $N$ binary spins $S_i=\pm1$ interact in triplets: 
\beq
\label{hdef}
H=-\rho N=-\sum_{i<j<k} C_{ijk} S_i S_j S_k 
\eeq
Here, the variable $C_{ijk}=1$ with $i<j<k$ denotes the presence 
of a plaquette connecting sites $i,j,k$, while $C_{ijk}=0$ 
denotes its absence.   
Choosing $C_{ijk}=1(0)$ randomly with probability 
$2c/N^2$ ($1-2c/N^2$) results in a random graph, where the number 
of plaquettes connected to a site is distributed with a 
Poisson distribution of average $c$ - this exemplifies
the {\it locally varying connectivities} mentioned above. The connection with granular compaction
is made in accordance with Edwards' 
hypothesis \cite{ergodic}: we interpret the
{\it local contribution to the energy in different 
configurations of the spins as the volume occupied by grains 
in different local orientations}. 

This Hamiltonian has
been studied on a random graph 
in various contexts \cite{hypersat, newman}. 
It has a trivial ground state where all spins 
point up and all plaquettes are in the configuration 
$+++$ giving a contribution of $-1$ to the energy. 
Yet, {\it locally}, plaquettes 
of the type $--+,-+-,+--$ (satisfied plaquettes) also give the same 
contribution, 
although one may not be able to cover the 
entire graph with these four types of plaquettes in equal 
proportions. 
This degeneracy of the four configurations of plaquettes with 
$s_i s_j s_k=1$ results in {\it frustration}, via the {\it competition} between satisfying 
plaquettes locally  and globally.
In the former case,
all states with even parity may be used, resulting in a large 
entropy and in the latter, only the $+++$ state may be used. 
Such frustration eventually leads to {\it slow dynamics}.

This mechanism has a suggestive analogy in the concept of geometrical 
frustration of granular matter, if we think of plaquettes
as granular clusters. When grains are shaken, 
they rearrange locally, but locally dense configurations can be mutually
incompatible. Voids could appear between densely packed clusters
due to mutually incompatible grain orientations between neighbouring
clusters. {\it The process of compaction in granular media consists
of a competition between the compaction of local clusters
and the minimisation of voids globally}~\footnote{ There are indications
\cite{jpcm,rods} that global minimisation of voids
can sometimes win out, when granular media are subjected to
 prolonged low-intensity vibration; this
results in a first-order jump to a 'crystalline' state over
significant length scales in the sample.}.

\begin{figure}[htb]
\begin{center}
\includegraphics[angle=90,width=.50\linewidth]{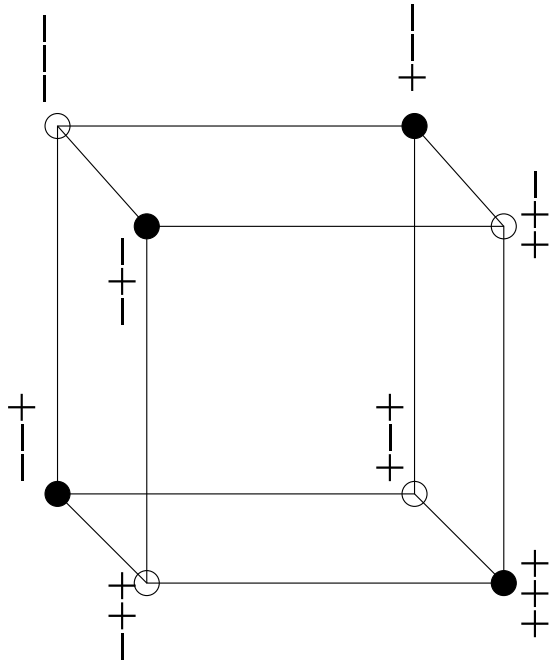}
\caption{\small
The phase space of three spins connected by a single plaquette. 
Configurations of energy $-1$ (the plaquette is satisfied) are indicated 
by a black dot, those of energy $+1$ (the plaquette is unsatisfied) are 
indicated by a white dot }
\label{figcube}
\end{center}
\end{figure}

Another key feature of this model is the existence of {\it metastable
states}. We note from Figure~\ref{figcube} that
{\it two} spin flips 
are required to take a given plaquette from one satisfied configuration 
to another; an energy barrier thus has to be crossed in any intermediate  
step between two satisfied configurations.
This has a mirror image
in the context of 
granular dynamics, where
compaction follows a temporary dilation; for example,
a grain could form an unstable ('loose') bridge with
other grains before it collapses into an available
void beneath the latter~\cite{jstat1, usbr}.
This mechanism, by which an energy 
barrier has to be crossed in going from one metastable state to 
another, is an important ingredient in models 
of granular compaction \cite{jpcm}.

\subsection{How we tap the spins - dilation and quench phases}

Our tapping algorithm
is a simplified version of the 
tapping dynamics used in cooperative Monte Carlo simulations
of sphere shaking \cite{usbr}. 
We treat each tap as consisting of two phases. First, during the {\it dilation}
phase, grains 
are provided with free volume to move into; next,
in the {\em quench} phase, they are allowed to relax until 
a mechanically stable configuration is reached. 

More technically,
the dilation phase is 
modelled by a single sequential Monte Carlo sweep of the system at a 
dimensionless temperature $\T$. A site $i$ is chosen at random and flipped with 
probability $1$ if its spin $s_i$ is antiparallel to its local field $h_i$, 
 with probability $\exp(-h_i/\T)$ if it is not, and with probability $0.5$ if 
$h_i=0$. This procedure is repeated $N$ times. 
Sites with a large absolute value of the local field $h_i$ 
thus have a low probability of flipping into the direction against the field; such spins may be thought of as being highly {\it constrained} by their neighbours. 
The dynamics of our 'thermal' dilation phase
 differs from the 'zero-temperature' dynamics used in 
\cite{dean} where a certain fraction of spins is 
flipped regardless of the value of their local field. 
Our choice \cite{johannes} reflects the following physics:
{\it if grains are densely packed ('strongly
bonded' to their neighbours), they are unlikely to be displaced
during the dilation phase of vibration}. 

The grains are then allowed to relax via a $\T=0$ quench, which 
lasts until the system has reached a {\it blocked} 
configuration~\cite{blocked}, 
where each site $i$ has $s_i=\mbox{sgn}(h_i)$ or $h_i=0$:
thus, each grain is either aligned with its local field, or
it is a 'rattler'~\cite{weeks}.
Thus, at the end of each tap (dilation $+$ quench), the 
system will be in a physically stable configuration~\cite{usbr}.

\subsection{The compaction curve}

\begin{figure}[htb]
\begin{center}
\includegraphics[angle=0,width=.50\linewidth]{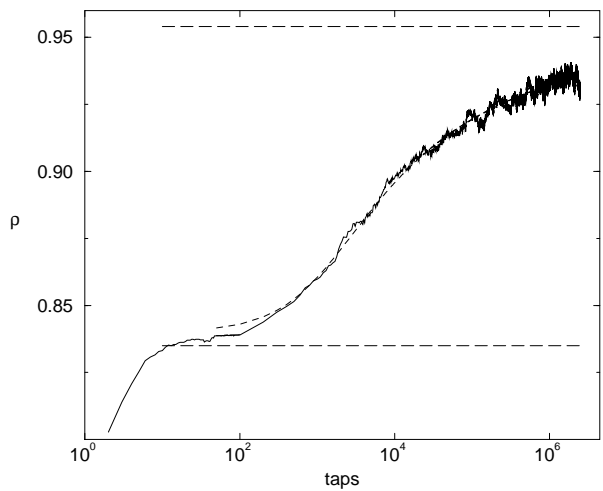}
\caption{\small
Compaction curve at connectivity $c=3$ for a system 
of $10^4$ spins (one spin is flipped at random per tap). 
The data stem from a single run with random 
initial conditions and the fit (dashed line) follows (\ref{loglaw}) 
with parameters $\rho_{\infty}=.971$, $\rho_0=.840$, $D=2.76$, 
and $\tau=1510$. The long-dashed line (top) indicates the approximate 
density $0.954$ at which the dynamical transition 
occurs, the long-dashed line (bottom) indicates the approximate 
density $0.835$ at which the fast dynamics stops, the 
{\it single-particle relaxation threshold}.}
\label{compact}
\end{center}
\end{figure}

Among the most important of our results is the compaction
curve obtained by tapping our model granular medium
 for long times. This is shown
in
Figure \ref{compact},
where three regimes of the dynamics can be identified. In the first
regime, {\it fast individual dynamics} predominates, while in the second, one sees
a logarithmic growth of the density via {\it slow collective} dynamics.The last
regime consists of {\it system-spanning density fluctuations} in the jamming limit, where our quantitative agreement with experiment~\cite{ed} allows us to
propose a {\it cascade theory of compaction during jamming}.

\subsubsection{Fast dynamics till SPRT: each grain for itself!}

At the end of
the first tap,
each grain is connected to 
more unfrustrated than frustrated clusters. This
is a direct result of the first tap being a zero-temperature
quench: any site where this was not the case
would simply flip its spin.
More generally, a {\it fast} 
dynamics occurs in this regime whereby 
{\it single} grains {\it locally} adopt the orientation
that, finally, optimises their density; this density
$\rho_0$ has been
termed~\cite{johannes} the
{\it single-particle relaxation threshold} (SPRT). 

If we neglect correlations between the local fields 
of neighbouring sites, we can arrive at a simple 
population dynamics model of the fast dynamics, details
of which can be found in~\cite{johannes}.
It yields a value of the SPRT,
$\rho_0=0.835$ (shown as a dotted line in Figure \ref{compact})
which is
{\it much higher} than the value of the density
$\rho(=0.49)$
of a typical 
blocked configuration. 

This is quite an extraordinary result; it implies that
despite the exponential dominance of
blocked configurations,
random initial conditions {\it preferentially select}
a higher density corresponding to the SPRT
$\rho_0$. This prediction of an overshoot in the
density achieved by fast dynamics has also, strikingly, been confirmed
in independent lattice-based models~\cite{usepl} of granular compaction,
and points to a strong {\it non-ergodicity} in the fast dynamics of individual grains. We will discuss this further in Section $5.3$.
 
Another significant feature of this regime is that
a fraction of spins is left with local fields exactly equal 
to zero, which thus keep changing orientation \cite{barratzecch}. 
These are manifestations in our model of 'rattlers' \cite{weeks}, 
i.e. grains which keep changing their orientation within well-defined 
clusters \cite{usbr}. They  will later (cf. Section $5.3$)  be used as a 
tool to probe the statistics of blocked configurations~\cite{johannes}.

To summarize: each grain
reaches its {\it locally} optimal configuration
via fast individual dynamics, resulting in the attainment
of the
SPRT density.
All dynamics
after this point is perforce collective.

\subsubsection{Slow dynamics of granular clusters: logarithmic compaction.}

The second stage of our tapping dynamics is {\it fully collective}: it
removes some of the remaining 
frustrated plaquettes as clusters slowly rearrange themselves.
A logarithmically slow compaction results
\cite{sid1, usepl},
leading from the SPRT density $\rho_0$ to the asymptotic density 
$\rho_\infty$. 
The resulting compaction curve may be fitted,
with $D$ and $\tau$ being characteristic constants,
 to the well-known logarithmic 
law \cite{sid1}:
\beq
\label{loglaw}
\rho(t)=\rho_{\infty}-(\rho_{\infty}-\rho_0)/(1+1/D \, \ln(1+t/\tau)) \ ,
\eeq 
This can be written more transparently as 
$1+t(\rho)/\tau = \exp{\{D \frac{\rho-\rho_0}{\rho_{\infty}-\rho} \}}\  $,
a form which makes clear that
the dynamics becomes slow (logarithmic) as soon as 
the density reaches $\rho_0$.  
Although most 
grains are firmly held in place by their neighbours
in this regime, {\it cascade-like} changes
of orientation can occur. For example, if some
grains change orientation 
during the dilation phase, this would change
the constraints on their neighbours; 
importantly,
the freer dynamics of rattlers could also alter local fields
in their neighbourhood,
and cause previously blocked grains to reorient.
Reorientation in cascades
\cite{johannes} would then ensue,
leading to collective granular compaction upto
the asymptotic density $\rho_\infty$.
We identify this with the density of random close packing~\cite{bernal} (RCP)
and
associate it with a {\it dynamical 
phase transition} \cite{monasson,franzpar}. 

\subsubsection{Cascades at the dynamical transition.}

 With increasing density, free-energy barriers rise up causing 
the dynamics to slow down according to (\ref{loglaw}). The point where 
the height of these barriers scales with the system size marks a
{\it breaking of the ergodicity of the dynamics};
an exponential number of valleys appears in the free-energy landscape.
Our theoretical~\cite{johannes} 
value for the dynamical transition, shown as
a horizontal line in Figure~\ref{compact}, is in good agreement
with the asymptotic density $\rho_\infty$ reached (numerically) by 
the tapping dynamics. 

\begin{figure}[htb]
\begin{center}
\includegraphics[angle=0,width=.50\linewidth]{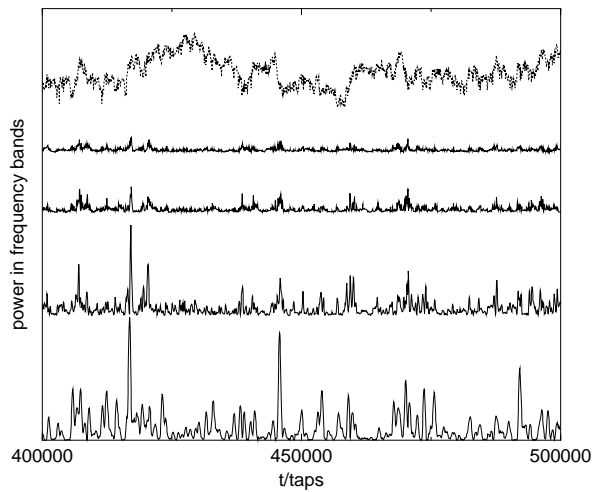}
\caption{\small
The density fluctuations as a function of time resulting from
1024 taps are plotted as the topmost trace.
The successive
plots are of the power spectrum against time,
in different frequency octaves.
The power in the 
first octave (frequency 1/(1024 taps) -2/(1024 taps))
is the bottommost trace, second octave 
(frequency 2/(1024 taps) - 4/(1024 taps)),
above it, and so on to the top.
Note that the fluctuations of the power in the 
different frequency bands are strongly correlated;
they 
correspond to sudden changes in the density (topmost trace).}
\label{freq}
\end{center}
\end{figure}

Also shown in  Figure~\ref{compact} are
marked fluctuations around the logarithmic 
compaction law, especially as the jamming limit
is approached; their correlations over several octaves
have been the subject of detailed experimental 
investigations \cite{ed}. To compare the results of our model with experiment,
 we follow 
\cite{ed} and use Fourier transforms to plot the power spectrum
of the timeseries $\rho(t)$ (in different frequency bands) against time 
in Figure \ref{freq}. 

Our results indicate, as in
the experimental data, that there
are 'bursts' in the power spectrum fluctuations: our decomposition
of these bursts over several octaves shows that they are caused
by strong correlations of
noise power over a wide range of 
frequencies. Importantly, the correlation
matrices we obtain~\cite{johannes} are in quantitative
agreement with experiment~\cite{ed}.

In our model, we can trace such bursts 
as being 
due to {\it cascades} of spin-flips. As mentioned before,
these arise from the change in local fields caused
by the flipping of a single spin (or several spins); this instability
propagates through ever larger neighbourhoods, causing 
correlated bursts in noise power fluctuations.
Note that this cascade mechanism is entirely absent from 
generic fully connected models~\cite{jorge}, where each spin interacts with 
all spins in the system. Interestingly, it is also absent from
the finitely-connected parking-lot model~\cite{plm}; this is because the
creation and filling of 'parking lots' by 'cars' in that model
 does not cause the
appearance of further parking lots, in the way spins flips 
may trigger a cascade. Put another way, the absence of {\it competition}
between individual and collective interactions in the  
parking-lot~\cite{plm} model results in an absence of cascades there, despite its finite connectivity.

We use the above insights to argue~\cite{johannes} that
in real granular media, the observed correlations~\cite{ed} of
density fluctuations are due to a {\it cascade} process in granular
compaction near the jamming limit. Here,
orientational/positional changes in strongly constrained
grains give rise to propagating instabilities, leading to a near-global
rearrangement of the granular medium. Pictorially, the movement
 of a single grain
in this regime is only possible as the consequence of a system-wide
cooperative motion of grains:
 this leads to sharp changes in overall density,
and to the observed 'bursts' in the power spectrum of density fluctuations~\cite{ed}.

\subsection{Realistic amplitude cycling; how granular media
jam at densities lower than close-packed.}

Our random graphs model has also been used to simulate {\it amplitude cycling},
an experimental~\cite{sid1} protocol on tapped granular media.
Here,
the granular medium
is tapped at a given amplitude $\T$ for a time $\tau$, after which
its amplitude is changed by an infinitesimal 
$\delta \T$; this process is repeated for cycles of increasing
and decreasing $\T$. The control parameter turns
out to be the so called 'ramp rate', which is
the ratio $\delta \T/\tau$; this is a measure~\cite{usepl} of the
'equilibration' allowed to the granular medium. Clearly low values
of this will correspond to quasi-static processes, while large
values will correspond to near-adiabaticity.

\begin{figure}[htb]
\begin{center}
\includegraphics[angle=0,width=.50\linewidth]{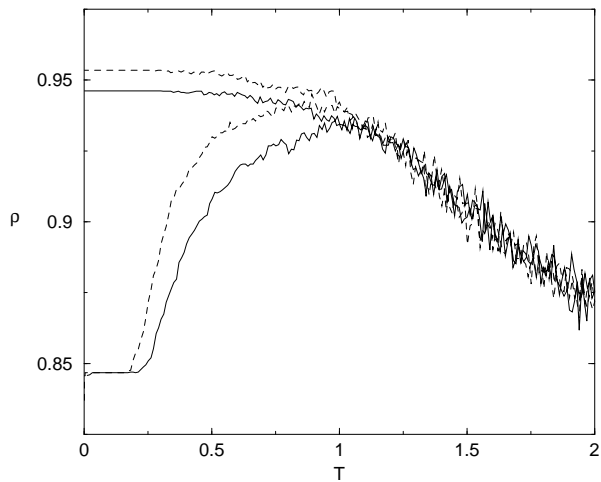}
\caption{\small
Amplitude cycling:
 $\T$ is varied in both directions between $0$ to $2$
at two rates $\delta \T = 10^{-4}, 10^{-5}$ per tap. with
$\tau = 1$. The lower ramp rate (shown by the dotted line)
 results in a higher final
density.
}
\label{figcycle1}
\end{center}
\end{figure}

The most simple-minded application of this protocol in our model
results in the scenario of 
Figure~\ref{figcycle1}.
At both high and low cycling rates, the 
density $\rho$ first reaches the SPRT $\rho_0$, increases 
with increasing amplitude, and decreases again at 
large values of $\T$. Thereafter, $\rho$ always decreases
with increasing $\T$. 
 The part of the curve where $\T$ is increased
 for the first time has been termed~\cite{sid1}
the {\it irreversible branch}; 
the {\it reversible branch} refers to the 
trajectory traced out by all successive increases and decreases of tapping
amplitude.
The results of Figure~\ref{figcycle1}, in agreement with many other
models 
\cite{usepl, jpcm, coniglio}, suggest
 that as the
ramp rate is decreased, the system will eventually attain
the 
RCP density 
$\rho_{\infty}$. In particular, these models
predict that
in the limit of near-zero ramp rates,
the irreversible branch disappears, with 
$\rho$ becoming a 
single-valued function of $\T$.

 This prediction is in direct
contradiction to the experimental results 
of \cite{sid1}; these suggest that 
at least for experimentally realisable times,
low-amplitude shaking does {\it not} result
in RCP being reached.
Instead, they suggest that
some grains in the jamming limit of
a granular 
assembly are so strongly constrained that they will {\it never} 
be displaced by low-amplitude taps.

In order to model the above experimental scenario, we modify the simplest
picture of amplitude cycling presented above. First, we realise
that in
the dynamics of our model described
thus far, sites with a high local field (corresponding to 
{\it strongly constrained}
grains) 
may be flipped at any finite value of $\T$ with a correspondingly 
{\it small but finite} probability. This is what eventually
 leads the system to the RCP density
$\rho_{\infty}$. 

To prevent this drift to RCP, one could naively think
of introducing a threshold
in local fields such that spins with fields above
this threshold are not flipped.
It turns out~\cite{johannes}, however, that 
this is insufficient; the orientational
dynamics of neighbouring spins will always
loosen the  constraints on previously blocked spins
 in the end,
and lower their local fields below any given threshold.
The above implies that the constraints
on grains are not related to {\it orientational} frustration alone;
it was suggested~\cite{johannes} that they might also be {\it mechanical}
in nature,
related to force networks~\cite{mueth} between grains. 
Further, it seemed reasonable to suppose that such
 effectively  immobile~\cite{bmmb} 
'blockages' could only be removed kinetically by the imposition of
large intensity ($\T$) vibrations.

Accordingly, the concept of
{\it low-amplitude pinning} of grains was introduced~\cite{johannes}:
 assign to each site $i$ a real number $r_i$
between zero and one, such that 
only grains with $r_i<\T$ (mechanically constraining forces
less than external vibration intensity) will be free to move.
This modification could in principle lead to a
lower value of $\rho_\infty$ after amplitude recycling: for example,
spin plaquettes (granular clusters) generated during
the high-amplitude part of the cycle, would be effectively
immobile at lower amplitudes, leading to wasted space. 

\begin{figure}[htb]
\begin{center}
\includegraphics[angle=0,width=.50\linewidth]{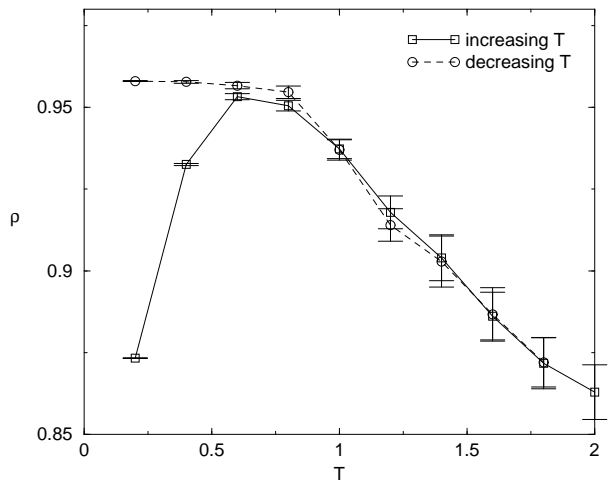}
\caption{\small
The asymptotic density for tapping amplitudes ranging from $\T=0.2$ to 
$\T=2$ in steps of $0.2$. The density measured after $10^7$ taps at each 
amplitude and convergence to a steady-state each time was checked.} 
\label{asympt} 
\end{center}
\end{figure}

However,
 this too is insufficient to stop the evolution of the system to
 $\rho_\infty$. It turns out~\cite{johannes} that
 jamming at lower densities can only be achieved if
low-amplitude pinning is {\it combined} with the choice of
an extremely low ramp rate, via a
large 'equilibration time' $\tau$. 
If this is done, grains are allowed to 'equilibrate'
 at each amplitude,   
thus making  
sure that a steady-state of the
 density is always reached.
This leads,
finally, to
the low-amplitude immobilisation  of granular clusters
created at high amplitudes, to the consequent blocking of voids,
and hence to 
'jamming' 
~\cite{sid1,nagel1}
at densities lower than RCP (see
Figure~\ref{asympt})

Our results~\cite{johannes} on modified
amplitude cycling thus indicate that
it is important to have {\it mechanical pinning as well as long equilibration times}~\cite{contrary} for jamming to occur.
In fact, the
random configurations of immobile spins at each value of 
$\T$ can be viewed~\cite{johannes} as an additional quenched disorder,
and their effect on neighbouring 
mobile spins, as an additional  random local field. 
Our results demonstrate also
 a rather fundamental difference between
excitations in glassy systems and
 granular media. In glasses, one would expect the configurations 
of spins reached at high values of temperature
and subsequently frozen, {\it immediately} to alter the behaviour of the system at 
lower values of temperature. In athermal media like
granular systems, however, it is important
to allow {\it equilibration} at each value of the shaking intensity
$\T$, in order even to begin to observe the hysteretic effects which result
in jamming at lower-than-RCP densities.

\subsection{Discussion}
In the above,
we have presented a
 {\it finitely connected} 
spin model on random graphs~\cite{johannes}, with
the aim of examining the compaction of tapped granular media.
Fast non-ergodic relaxation of individual grains terminated
at the SPRT density, in our compaction curve; collective relaxation
followed, manifest first by logarithmic compaction, and next by
system-wide density fluctuations around RCP, both of which match
experimental results~\cite{sid1, ed}. Our model explains the latter
in terms of a {\it cascade} process that occurs near the jamming limit
of granular matter. Also, in contradistinction to other models~\cite{usepl, 
jpcm, coniglio}, our results indicate that jamming at densities
lower than RCP occurs as a result of {\it competition between mechanical
and orientational frustration}, during amplitude cycling. We leave
the discussion of the configurational entropies~\cite{johannes, jproc}
 generated by our tapping
 algorithm to the concluding section of this article.

\section {Shape matters}

In the concluding section of this review, we explore the effect
of {\it grain shapes}
in granular compaction.
Our model is based on the following picture. Consider a box
of sand in the presence of gravity; for ease of visualisation,
we think of this as being constituted of vertical columns.
In the jamming limit, diffusion of grains {\it between} columns is 
inhibited~\cite{usepl}
due to the absence of holes (grain-sized voids); compaction proceeds
instead by grain reorientation {\it within} each column~\cite{column}
 to minimise the size of the partial
voids that persist.
This intuitive picture is verified by results of
computer simulations~\cite{usbr} which show that correlations
in the transverse (inter-column) direction are negligible
compared to those in the longitudinal (intra-column) direction.
We thus focus on a column model of grains in the jamming limit~\cite{column}.

 Each ordered grain occupies one unit of space,
while each disordered grain occupies $1+\eps$ units of space, with
$\eps$  a measure of the {\it partial void trapped by misorientation}.
A reorientation of the grain to an ordered (`space-saving') state frees up
the partial void, for use by other grains to reorient themselves.
This cascade-like picture of compaction
 in the jamming limit  resembles that presented
in the previous section, where 
random graph models~\cite{johannes} were discussed. 
Also, as there, the response of
grains to external dynamics is via the {\it local} minimisation of void space
modelled by a local field.

\subsection{A model of regular and irregular grains}
In our column model,
grains are indexed by their depth $n$ measured from the free surface.
Each grain can be in one of two orientational states --
ordered ($+$) or disordered ($-$) --
the `spin' variables $\{\s_n=\pm1\}$ thus uniquely defining a configuration.
Exactly as in the random graphs model~\cite{johannes} presented above,
a local field $h_n$ 
constrains the temporal evolution of spin $\s_n$,
such that excess void space is minimised.

In the presence of a vibration intensity $\T$,
grains reorient with an ease that depends on
their depth $n$ within the column (grains at the free surface
must clearly be the freest to move!), as well as on the local
void space $ h_n$ available to them. The transition
probabilities governing this are:
\beq
w\bigl(\s_n=\pm\to\s_n=\mp\bigr)=\exp\bigl(-n/\xidy\mp h_n/\T\bigr).
\label{w}
\eeq
The dynamical length $\xidy$~\cite{column, usepl}
 effectively defines the boundary layer 
of the column; within this dynamics are {\it fast}, while
well beyond it, they are {\it slow}.
The local field $h_n$ is a measure of 
{\it excess void space}~\cite{br}: 
\beq
h_n=\eps\,m^-_n-m^+_n,
\label{ydef}
\eeq
where $m^+_n$ and $m^-_n$ are respectively the numbers of $+$
and $-$ grains above grain $n$. The definition
Equation~(\ref{ydef}) is such that a transition
from an ordered to a disordered state for grain $n$ is
{\it hindered} by the number of voids that are already above it, as might
be expected for an ordering field in the jamming limit.

In the $\T\to0$ limit of zero-temperature dynamics~\cite{johannes},
the probabilistic rules~(\ref{w}) become deterministic: the expression
$\s_n=\sign\,h_n$ 
(provided $h_n\ne0$)
 determines the {\it ground states} of the system.
{\it Frustration}~\cite{spinglass} manifests itself for
 $\eps>0$, which leads to a rich ground-state structure,
whose precise nature depends on whether $\eps$ is rational or irrational.
We mention for completeness that the case $\eps<0$
discussed in earlier work \cite{usepl}
corresponds to a {\it complete absence of frustration and a single
ground state of ordered grains}.

For irrational $\eps$, no local field $h_n$ can ever be zero 
(cf.~(\ref{ydef})).
Noting that
 irrational values of $\eps$ denote shape irregularity,
we conclude that
the {\it excess void space
is nonzero even in the ground state of jagged grains}.
Their ground state, far from being perfectly packed, turns out~\cite{column} 
to be
quasiperiodic.

Regularly shaped grains correspond to
 rational $\eps=p/q$, with $p$ and $q$ mutual primes. We see
from~(\ref{ydef}) that now, 
 some of the $h_n$ can vanish; these correspond, as noted in the previous
section, to
 rattlers. 
A rattler at depth $n$ thus has a perfectly packed column above it,
so that it is free to choose its orientation~\cite{johannes, column, weeks}.
For regular grains in their ground state, rattlers occur
periodically (as in crystalline packings!) at points such that
$n$ is a multiple of the period $p+q$.\footnote{For example, when $\eps=1/2$,
each disordered grain `carries' a void half its
size; units of perfect packing must be permutations
of the triad $+--$, where two `half' voids from each
of disordered grains are perfectly filled by an ordered grain.
The {\it stepwise compacting} dynamics~\cite{column} selects only two of these
patterns, $+--$ and $-+-$}.
Every ground state is thus a random sequence of two patterns of length $p+q$,
each containing $p$ ordered and~$q$ disordered grains; this degeneracy
leads to 
a {\it zero-temperature configurational entropy}
or {\it ground-state entropy} $\Sigma=\ln 2/(p+q)$ per grain.

\subsection{Zero-temperature dynamics: (ir)retrievability of ground states, 
density fluctuations and anticorrelations}

Regular and irregular grains behave rather differently
when submitted to zero-temperature dynamics. The (imperfect) but unique
ground state for irregular grains is rapidly retrieved;
the perfect (and degenerate) ground states for regular grains never are,
resulting in {\it density fluctuations}.

We recall the rule for zero-temperature dynamics:
\beq
\s_n\to\sign\h_n.
\label{zerody}
\eeq
Starting with  irregular grains (with a given irrational value of
 $\eps$) in  
an initially disordered state,
one quickly recovers the ground state with zero-temperature
dynamics.
The ground state in fact propagates ballistically from the free surface
to a depth $L(t)\approx V(\eps)\,t$~\cite{column} at time $t$, while
the rest of the system remains in its disordered initial state.
When $L(t)$ becomes comparable with $\xidy$,
the effects of the free surface begin to be damped.
In particular for $t\gg\xidy/V(\eps)$
we recover the logarithmic coarsening law $L(t)\approx\xidy\ln t$,
also seen in other theoretical models~\cite{johannes, usepl}
of the slow relaxation of tapped granular media~\cite{sid1}. To recapitulate,
{\it the ground state for irregular grains is quickly (ballistically) recovered
with zero-temperature dynamics,
until the boundary layer
  $\xidy$ is reached; below this, the column is essentially
frozen, and coarsens only logarithmically}.

For regular grains with rational $\eps$, the local field $h_n$
in~(\ref{zerody}) vanishes for rattlers. Their dynamics
is stochastic even at zero temperature, since they have a choice
of orientations: a simple way to update them is according to the rule
$\s_n\to\pm1$ with probability $1/2$.
This stochasticity results in an intriguing dynamics
 even well within the boundary layer
$\xidy$, while the dynamics for $n\gg\xidy$
is as before logarithmically slow~\cite{column}.

In what follows, we will focus on the fast dynamics within
the boundary layer. Our
main result is that zero-temperature dynamics
does not drive the system to any of its degenerate ground states,
but instead engenders a {\it fast relaxation to a non-trivial steady state},
independent of initial conditions, which  consists of
 {\it unbounded density fluctuations}.
This recalls density fluctuations
close to the jamming limit~\cite{johannes, sid1}, in other studies
of granular compaction.

\begin{figure}[htb]
\begin{center}
\includegraphics[angle=90,width=.50\linewidth]{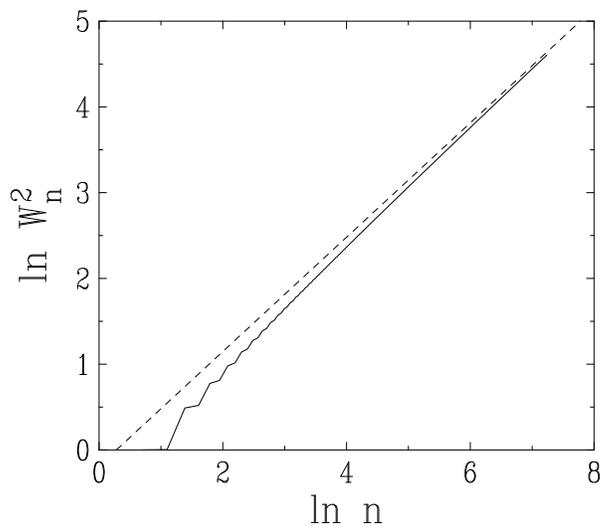}
\caption{\small
Log-log plot of $W_n^2=\mean{h_n^2}$ against depth $n$,
for zero-temperature dynamics with $\eps=1$.
Full line: numerical data.
Dashed line: fit to asymptotic behaviour leading to~(\ref{rough})
(after~\cite{column}).}
\label{fign}
\end{center}
\end{figure}

\begin{figure}[htb]
\begin{center}
\includegraphics[angle=90,width=.50\linewidth]{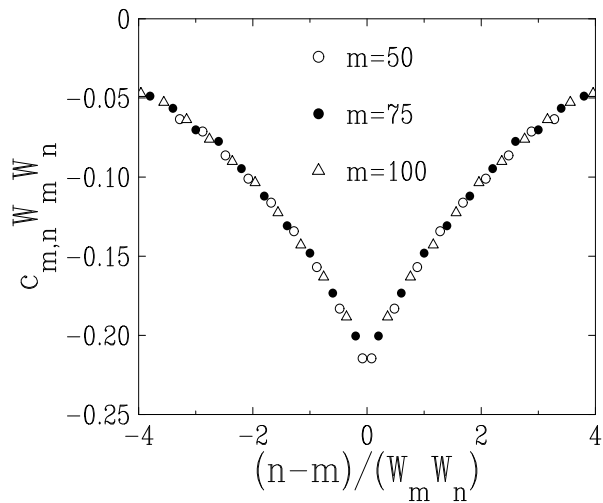}
\caption{\small
Scaling plot of the orientation correlation function $c_{m,n}$ for $n\ne m$
in the zero-temperature steady state with $\eps=1$,
demonstrating the validity of~(\ref{cmn})
and showing a plot of (minus) the scaling function $F$
(after~\cite{column}).}
\label{figo}
\end{center}
\end{figure}

Fig.~\ref{fign} shows the variation of these
density fluctuations as a function of depth $n$:
\beq
W_n^2=\mean{h_n^2}\approx A\,n^{2/3},\quad A\approx 0.83.
\label{rough}
\eeq
The fluctuations are approximately Gaussian,
with a definite excess at {\it small} values:
$\abs{h_n}\sim 1\ll W_n$. We recall that 
non-Gaussianness was also observed in
experiments on density
fluctuations in tapped granular media~\cite{ed}; in our theory here,
we interpret it in terms of grain (anti)correlations.
If grain orientations were fully uncorrelated,
one would have the simple result $\mean{h_n^2}=n\eps$,
while~(\ref{rough})  implies
that $\mean{h_n^2}$ grows much more slowly than~$n$.

It turns out that at least within a  dynamical cluster of radius 
$n^{2/3}$)~\cite{column},
the orientational displacements of each grain are {\it fully
anticorrelated}. Fig.~\ref{figo} shows that
the orientation correlations $c_{m,n}=\mean{\s_m\s_n}$
scale as~\cite{column}
\beq
c_{m,n}\approx\delta_{m,n}
-\frac{1}{W_mW_n}\,F\!\left(\frac{n-m}{W_mW_n}\right),
\label{cmn}
\eeq
where the function $F$ is such that
$\int_{-\infty}^{+\infty}F(x)\,{\rm d}x=1$.
We find also that, within such a dynamical cluster,
the fluctuations of the orientational displacements are 
 {\it totally screened}:
$\sum_{n\ne m}c_{m,n}\approx-c_{m,m}=-1$.
These results recall the {\it anticorrelations
in grain displacements} observed in independent
simulations of shaken hard spheres close to jamming~\cite{usbr};
there they corresponded to compaction via bridge collapse,
as upper and lower grains in bridges~\cite{br} collapsed onto
each other, releasing void space.

\subsection{Rugged entropic landscapes: Edwards' or not?}

The most remarkable feature of our column model is, arguably, the
rugged landscape of microscopic configurations visited 
during the steady state of zero-temperature dynamics (for regular
grains);
this is all the more
striking because the macroscopic entropy is {\it flat}, in agreement
with Edwards' hypothesis~\cite{ergodic}. 

\begin{figure}[htb]
\begin{center}
\includegraphics[angle=90,width=.5\linewidth]{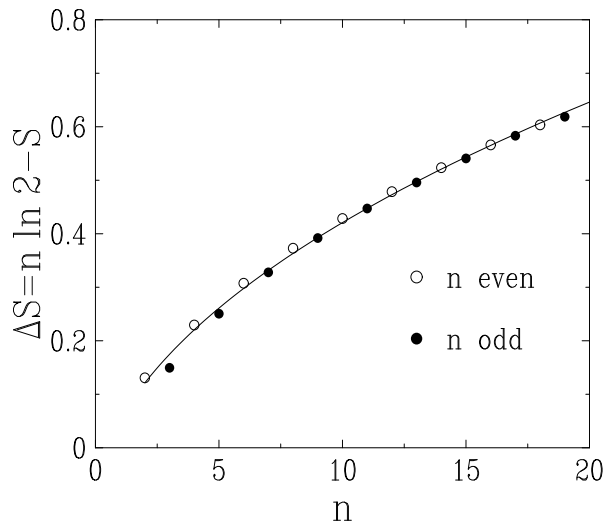}
\caption{\small
Plot of the measured entropy reduction $\Delta S$
in the zero-temperature steady state with $\eps=1$,
 against $n\le19$.
Symbols: numerical data,  for $t\sim10^9$ and $n\approx20$.
Full line: fit $\Delta S=(62\ln n+53)10^{-3}\,n^{1/3}$.}
\label{figh}
\end{center}
\end{figure}

The entropy of the steady state of zero-temperature dynamics  is 
defined by the usual Boltzmann formula
\beq
S=-\sum_\C p(\C)\ln p(\C),
\label{boltz}
\eeq
where $p(\C)$ is the probability that the system is
in the orientation configuration $\C$ in the steady state,
and the sum runs over all the $2^n$ configurations
of a system of $n$ grains.
This can be estimated theoretically  
by using~(\ref{rough}).
Consider $n$ as a fictitious discrete time, with
the local field $\h_n$ as the position of a random walker at time $n$.
For a free lattice random walk of $n$ steps,
one has $\mean{h_n^2}=n$;
as all configurations are equiprobable.
 the entropy reads $S_\flat=n\ln 2$. For a column of
regularly shaped grains,
our model predicts instead $\mean{\h_n^2}=W_n^2\ll n$; the
 entropy $S$ of our random walker is therefore
 reduced with respect to $S_\flat$.
The entropy reduction~\cite{remi} $\Delta S=S_\flat-S=n\ln 2-S$
can be estimated~\cite{column} to be
\beq
\Delta S\sim\sum_{m=1}^n\frac{1}{W_m^2}\sim n^{1/3}.
\label{stheo}
\eeq

\begin{figure}[htb]
\begin{center}
\includegraphics[angle=90,width=.5\linewidth]{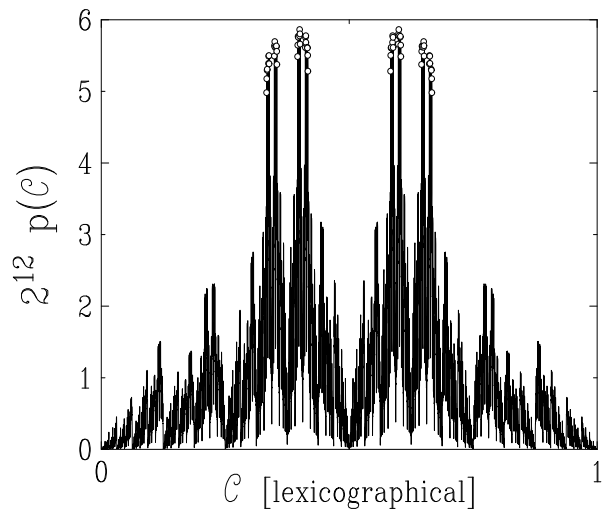}
\caption{\small
Plot of the normalised probabilities $2^{12}\,p(\C)$ of the configurations
of a column of 12 grains in the zero-temperature steady state with $\eps=1$,
against the configurations $\C$ in lexicographical order.
The empty circles mark the $2^6=64$ ground-state configurations,
which turn out to be the most probable (after~\cite{column}).}
\label{figigs}
\end{center}
\end{figure}

Evaluating the steady-state entropy $S$ numerically,
using~(\ref{boltz}) and measuring all configurational
 probabilities $p(\C)$, we find (cf. Figure~\ref{figh})
that
 $\Delta S$ is small; for example for
$n=12$, we have $\Delta S\approx0.479$, in good agreement
with the results of Eq.~(\ref{stheo}). 
This is a convincing demonstration 
 that anticorrelations (see previous section)
lead to only microscopic corrections to the overall dynamical
entropy of the steady state, which is
{\it flat}, in agreement with Edwards' hypothesis~\cite{ergodic}.

To investigate the effect of the constraints, we plot the
normalised configurational probabilities $2^{12}\,p(\C)$ for
a column of 12 grains against
the $2^{12}=4096$ configurations $\C$ in
Figure~\ref{figigs}.
Note that the actual values of the configurational probabilities
$p(\C)$ are microscopically small! At this microscopic scale,
however, the entropic landscape is
startlingly rugged;
some configurations are clearly visited far more often than others.
It turns out that the most visited configurations
are the ground states of the system (empty circles).
We suggest that this behaviour is generic: i.e.,
{\it the dynamics of compaction in the jammed state leads to a microscopic
sampling of configuration space which is highly non-uniform, so that
its ground states are visited most frequently}. 
Our model thus provides a natural reconciliation between, on the one hand,
the intuitive perception that not all configurations can be equally visited
during compaction in the jamming limit, that the most compact
configurations should  be the most visited; and, on the other,
the flatness hypothesis
of Edwards, which states that for large enough systems, the entropic
landscape of visited configurations is flat~\cite{ergodic}.

The dynamical entropy generated by the random graphs model~\cite{johannes} of
the previous section is also reconcilable with Edwards' 
flatness~\cite{ergodic},
at least in the jamming limit discussed above. This was explored
via rattlers ( sites $i$ such that the local
field $h_i=0$) in the blocked configurations generated 
after each tap. We have seen above that
they have a rather crucial role to play the density fluctuations of our 
column model~\cite{column}; it turns out
that they are also a good probe of Edwards' flatness under
the tapping dynamics of our random graphs model. If
blocked 
states at a given density are equiprobable, a plot 
of the fraction of connected rattlers versus the density 
should reproduce this~\cite{johannes}.

\begin{figure}[htb]
\begin{center}
\includegraphics[angle=0,width=.50\linewidth]{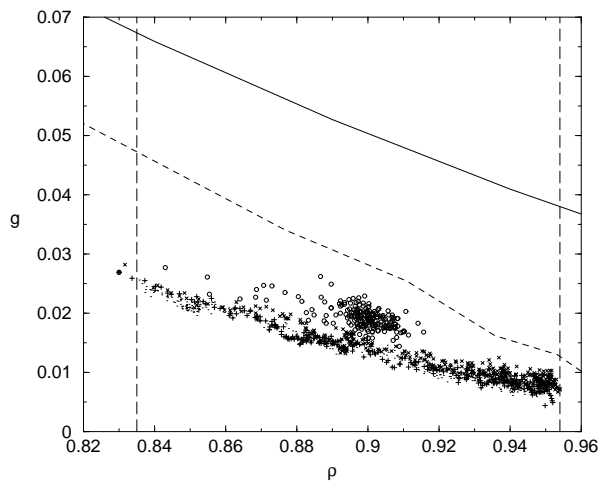}
\caption{\small
The fraction $g$ of connected rattlers 
during $4$ runs of a tapped random graph~\cite{johannes} with $N=1000, c=3$ at $T=0.4$ (dots), 
$T=0.56$ ($+$), $T=0.7$ ($\times$), and $T=1.5$ (circles).
The solid and dashed lines correspond respectively to annealed
and quenched theoretical values corresponding to Edwards' flatness.
The vertical lines
indicate the approximate values for $\rho_0$ (left line) and $\rho_{\infty}$ (right line) 
respectively.}
\label{figedwards_test}   
\end{center}
\end{figure}

Figure \ref{figedwards_test} shows the results for four single runs
of plotting the fraction of rattlers 
 $g$ against density $\rho$, 
  at increasing amplitudes of vibration $\T$.
The dashed line and full lines correspond respectively
to quenched and annealed replica symmetric averages
for $g$, assuming Edwards' flatness~\cite{johannes}.
 We notice that there is a reasonable congruence of all
the numerical results and the (theoretically more accurate) quenched
average at the asymptotic density $\rho_{\infty}$. Thereafter, there are systematic divergences with lower density and higher $\T$. 

We can draw the following conclusions from this. First, at the jamming
limit near RCP, the dynamically generated entropies are flat, in accord
with Edwards' hypothesis~\cite{ergodic}, as well as with
the results of our column model~\cite{column}. 
Second, as we move to the regimes
of higher vibration and lower density, the entropic landscape
gets rougher - one can imagine a process whereby
the roughening visible on microscopic scales near jamming (cf. Figure~\ref{figigs}) begins to increase to macroscopic scales as one moves away from jamming.
In this regime, we observe that configurations which are dynamically
accessed by tapping (cf. the symbols in Figure~\ref{figedwards_test})
 correspond to {\it higher than typical densities} 
(dashed and full lines in Figure~\ref{figedwards_test}) - we recall from 
section 4.3.1 that this
occurs when 
 non-ergodic fast dynamics dominates configurational access. Putting
all of this together, we conclude that also according to our random
graphs model, Edwards' flatness~\cite{ergodic} governs dynamically generated configurational
entropies when slow (ergodic) dynamics predominate in the jamming limit; there are however, systematic deviations from flatness when
fast dynamics predominate, in the regime of higher tapping amplitudes and lower densities.

\begin{figure}[htb]
\begin{center}
\includegraphics[angle=0,width=.35\linewidth]{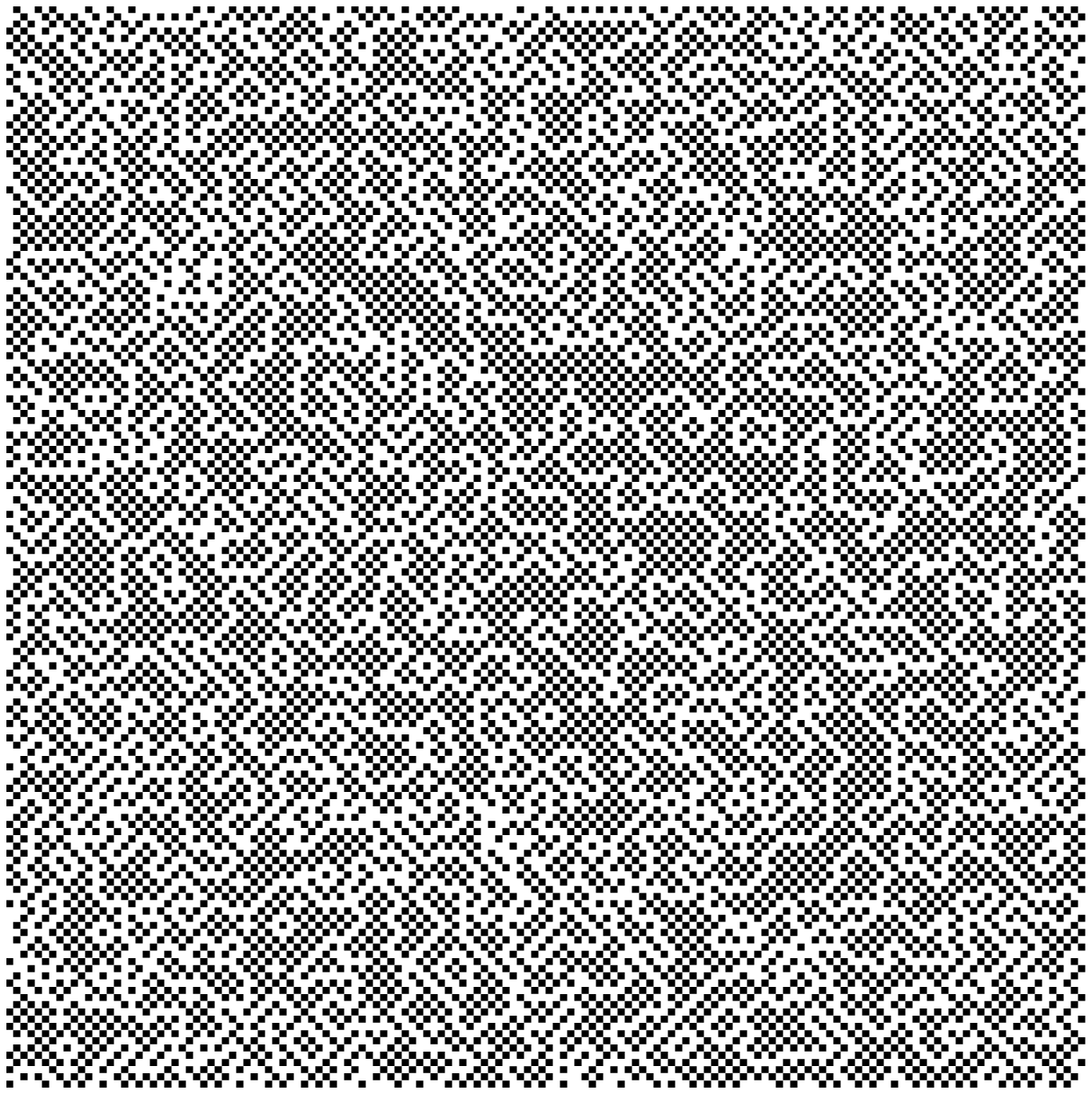}
{\hskip 1mm}
\includegraphics[angle=0,width=.35\linewidth]{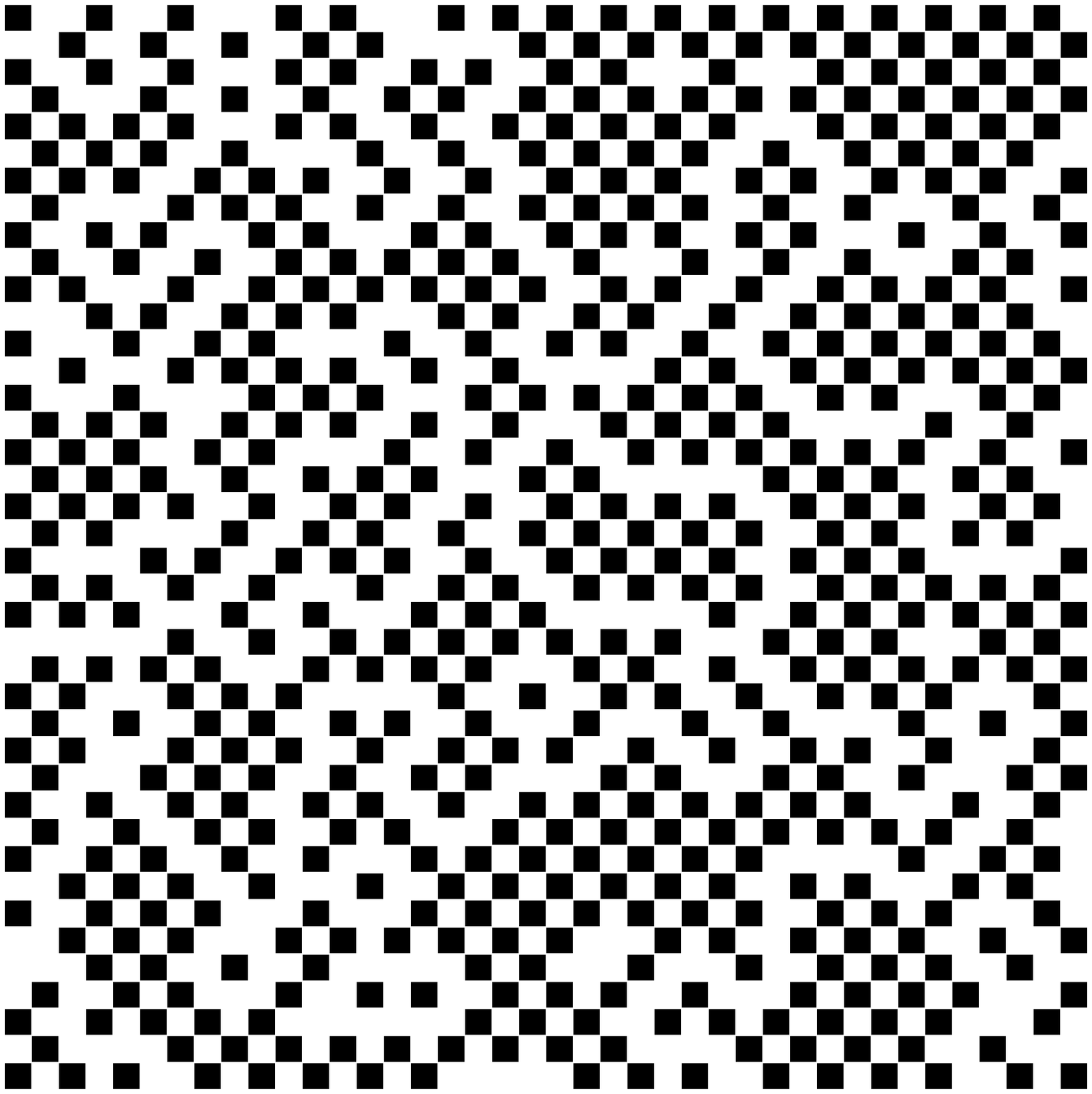}
\caption{\small 
A typical pattern of surviving clusters on the square lattice
for the cluster aggregation model of Refs.~\cite{condmat1, condmat2}.
Black (resp.~white) squares represent $\s_\n=1$ (resp.~$\s_\n=0$),
i.e., surviving (resp.~dead) sites. The left panel shows a
$150^2$ sample, while the right panel is enlarged ($40^2$) for clarity.}
\label{fige}
\end{center}
\end{figure}

Configurational entropies of strongly nonequilibrium models with slow dynamics, are, however,
not generically flat. To demonstrate this,
we present results for a model of nonequilibrium
 aggregation, which despite its
origins in cosmology~\cite{archan} turns out to have applications
in the gelation of stirred colloidal solutions~\cite{condmat1, condmat2}. This
  'winner-takes-all' model of cluster growth, whereby the largest cluster always wins, manifests both fast and slow dynamics. In mean field, the slow dynamical phase results
in at most one surviving cluster at asymptotic times; however, on finite
lattices, there can be many {\it metastable} clusters which survive forever, 
provided they
are each isolated from the others (Figure~\ref{fige}).

\begin{figure}[htb]
\begin{center}
\includegraphics[angle=90,width=.6\linewidth]{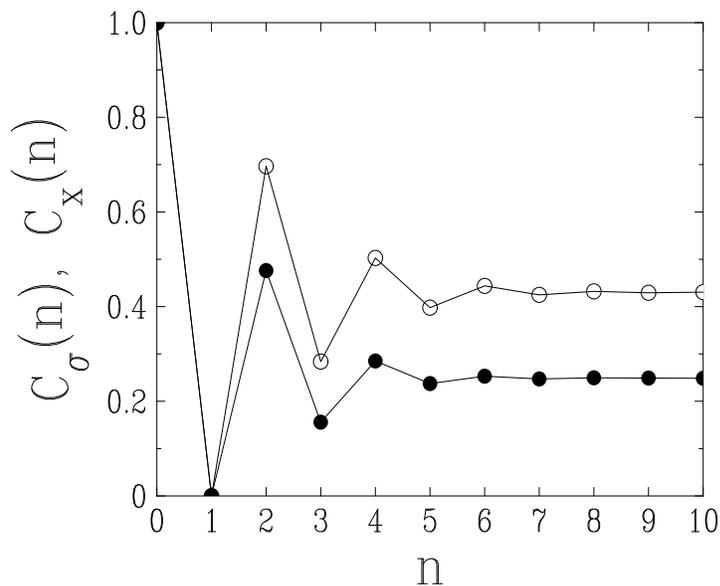}
\caption{\small Plot of correlation functions for the cluster
aggregation model of Refs.~\cite{condmat1, condmat2}
against the distance $n$ along the chain.
Empty symbols: correlation $C_\s(n)$ of the survival index.
Full symbols: correlation $C_x(n)$ of the reduced mass.}
\label{figf}
\end{center}
\end{figure}

We remark that this 'isolation' of surviving sites implies a very
strong anticorrelation between neighbouring sites in this model; that is,
each survivor must have voids around it, or run the risk of dying out. These
anticorrelations are manifest in Figure~\ref{figf}, both for cluster
survival and cluster mass on a one-dimensional version
of the model. The presence of such anticorrelations 
and of competition between slow and fast dynamics in a nonequilibrium context,
 suggests strong analogies between this
model~\cite{condmat1, condmat2} and our random graphs~\cite{johannes} and
column~\cite{column} models. We might therefore naively expect some version
of Edwards' flatness to hold; however, our results~\cite{condmat1} suggest
that it does {\it not}.

In conclusion, we emphasise that Edwards' flatness in the landscape
of configurational entropies is {\it not} the generic fate of strongly nonequilibrium
models with slow dynamics, even when they have many features in common. The similarity between our column model~\cite{column} and the random graphs model~\cite{johannes} discussed in the previous section is thus all the more remarkable;
both models manifest Edwards' flatness in the jamming limit, deviating
from it whenever free volume constraints are relaxed.

\subsection{Low-temperature dynamics along the column: intermittency}

We now return to the investigation of the column model~\cite{column},
 to do with its
 low-temperature dynamics. For rational $\eps$, the presence of a finite but low shaking intensity
 merely increases the magnitude of density fluctuations~\cite{sid1}, given
that the zero-temperature dynamics is
 in any
case stochastic. However, for irrational $\eps$, low-temperature dynamics
introduces an {\it intermittency in the position of a surface layer};
this has recently been observed in
experiments on vibrated granular beds~\cite{eric}.

This happens as follows: when the shaking amplitude 
$\T$ is such that it does not distinguish between a very small
void $h_n$ and the strict absence of one, the site $n$ `looks like'
a point of perfect packing. The grain at depth $n$ then
has the freedom to point the `wrong' way;
we call such sites {\it excitations}, using the thermal analogy.
The probability of observing an excitation at site $n$ scales as
$\Pi(n)\approx\exp(-2\abs{h_n}/\T)$.
The uppermost site $n$ such that
 $\abs{h_n}\sim\T\ll1$ will be the 'preferred' excitation; it
 is propagated ballistically
(cf. zero-temperature irrational $\eps$ dynamics)
until another excitation is nucleated above it. Its instantaneous position
$\N(t)$ denotes the layer at which shape effects are lost in thermal noise,
i.e., it separates an upper region of quasiperiodic
ordering from a lower region of density fluctuations~(\ref{rough}).

\begin{figure}[htb]
\begin{center}
\includegraphics[angle=90,width=.50\linewidth]{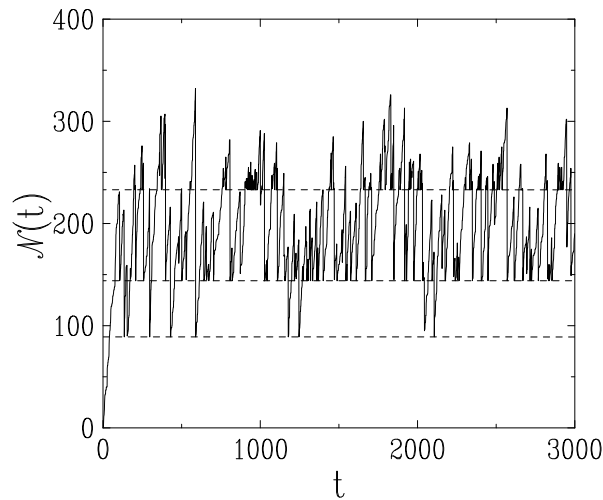}
\caption{\small
Plot of the instantaneous depth $\N(t)$ of the ordered layer,
for $\eps=\Phi$ (the golden mean) and $\T=0.003$.
Dashed lines: leading nucleation sites given by Fibonacci numbers
(bottom to top: $F_{11}=89$, $F_{12}=144$, $F_{13}=233$) (after~\cite{column}).}
\label{figk}
\end{center}
\end{figure}

Fig.~\ref{figk} shows a typical sawtooth plot of the instantaneous
depth $\N(t)$, for a temperature $\T=0.003$.
The {\it ordering length}, defined as $\mean{\N}$,
is expected to diverge at low temperature,
as excitations become more and more rare; we find in fact~\cite{column}
a divergence of the ordering length at low temperature of the form
$\mean{\N}\sim1/(\T\abs{\ln\T})$.
This length is a kind of finite-temperature equivalent of the
`zero-temperature' length $\xidy$, as it divides an ordered
boundary layer from a lower (bulk) disordered region.
We emphasise once again that fast dynamics predominates
in
both zero-temperature and finite temperature
boundary layers ($\xidy$ and $\mean{\N}$), with slow dynamics
setting in for column depths beyond these.

\subsection{Discussion}
We have discussed the effect of shape in granular compaction
near the jamming limit, via a column model of grains~\cite{column}. Our
main conclusions are that
jagged (irregular) grains are 
characterised by optimal ground states, which are easily
retrievable, while smooth (regular) grains cannot retrieve their ground states
of perfect packing; in the latter case, even zero-temperature dynamics
results in density fluctuations. We predict also that grain
irregularities result in a surface intermittency for low-amplitude shaking,
in agreement with recent experimental observations~\cite{eric}.

\section{Conclusions}
We have here reviewed some of our approaches to 
granular dynamics, now well known to consist of both
fast and slow relaxational processes~\cite{book}.
In the first case, grains typically compete
with each other, while in the second, they cooperate. 
The dynamics of bridge formation is a typical result of
 cooperation, as is the relaxation of the angle of repose; competition
between density fluctuations
 and external
driving forces can, on the other hand, result in sandpile collapse.
In the random graphs model presented above,
the SPRT density separates regions of cooperation and competition, each
with its own distinctive features. Finally,
while slow dynamics predominate deep inside our column model of
compacting grains, fast dynamics gives rise
to strikingly rough configurational landscapes and surface intermittency.

\end{document}